%% file: paper.tex
\documentclass[12pt]{iopart}

\usepackage{color}
\usepackage{epsfig}
\usepackage{float}
\usepackage{multirow}
\expandafter\let\csname equation*\endcsname\relax
\expandafter\let\csname endequation*\endcsname\relax
\usepackage{amsmath}

\begin{document}


\title[The characterization of Virgo data and its impact on gravitational-wave searches.]
{The characterization of Virgo data and its impact on gravitational-wave searches.}

\input{authors_list}


\ead{robinet@lal.in2p3.fr}

\input{abstract}

\maketitle

\input{introduction}

\input{virgo}

\input{detchar}
\input{glitch}

\input{lines}

\input{analyses}

\input{conclusion}

\input{acknowledgments}

\bibliographystyle{iopart-num}

\section*{References}
\bibliography{references}

\end{document}

%% file: authors_list.tex
\author{J.~Aasi$^{1}$, 
J.~Abadie$^{1}$, 
B.~P.~Abbott$^{1}$, 
R.~Abbott$^{1}$, 
T.~D.~Abbott$^{2}$, 
M.~Abernathy$^{3}$, 
T.~Accadia$^{4}$, 
F.~Acernese$^{5ac}$, 
C.~Adams$^{6}$, 
T.~Adams$^{7}$
% ...
}
\address{$^{1}$LIGO - California Institute of Technology, Pasadena, CA  91125, USA }
\address{$^{2}$California State University Fullerton, Fullerton CA 92831 USA}
\address{$^{3}$SUPA, University of Glasgow, Glasgow, G12 8QQ, United Kingdom }
\address{$^{4}$Laboratoire d'Annecy-le-Vieux de Physique des Particules (LAPP), Universit\'e de Savoie, CNRS/IN2P3, F-74941 Annecy-Le-Vieux, France}
\address{$^{5}$INFN, Sezione di Napoli $^a$; Universit\`a di Napoli 'Federico II'$^b$, Complesso Universitario di Monte S.Angelo, I-80126 Napoli; Universit\`a di Salerno, Fisciano, I-84084 Salerno$^c$, Italy}
\address{$^{6}$LIGO - Livingston Observatory, Livingston, LA  70754, USA }
\address{$^{7}$Cardiff University, Cardiff, CF24 3AA, United Kingdom }
% ...

%% file: abstract.tex
%

\begin{abstract}

Between 2007 and 2010 Virgo collected data in coincidence with the
LIGO and GEO gravitational-wave (GW) detectors. These data have been searched
for GWs emitted by cataclysmic phenomena in the universe, by
non-axisymmetric rotating neutron stars or from a stochastic
background in the frequency band of the detectors. The sensitivity of
GW searches is limited by noise produced by the detector or its
environment. It is therefore crucial to characterize the various
noise sources in a GW detector. This paper reviews the Virgo detector noise 
sources, noise propagation, and conversion mechanisms which were
identified in the three first Virgo observing runs. In many cases,
these investigations allowed us to mitigate noise sources in the
detector, or to selectively flag noise events and discard them from
the data. We present examples from the joint LIGO-GEO-Virgo GW
searches to show how well noise transients and narrow spectral lines
have been identified and excluded from the Virgo data. We also discuss
how detector characterization can improve the astrophysical reach of
gravitational-wave searches.
\end{abstract}

%% file: introduction.tex
%

\section{Motivations}\label{sec:introduction}

The first-generation gravitational wave (GW) interferometric
detectors, TAMA300~\cite{1742-6596-120-3-032010}, LIGO~\cite{0034-4885-72-7-076901},
GEO600~\cite{0264-9381-27-8-084003} and Virgo~\cite{The_Virgo_Detector_paper}, have performed
several GW searches over the last decade. In 2007-2010, LIGO,
GEO600 and Virgo detectors operated in coincidence at, or near, their
design sensitivities. Many noise events and non-stationarities are
present on top of the fundamental Gaussian component of the detector
output, and so searches for GW events require signals to be observed
in multiple detectors to reduce the large number of false-alarm events
due to instrumental or environmental disturbances. A multi-detector
network also offers a better sky coverage and the possibility of
localizing the source's sky position. The detection of a GW event is
expected to be unlikely given the detector sensitivities in 2007-10,
and, in the analyses performed so far, no GW signal has been detected
by the LIGO-GEO-Virgo network of interferometers.

In GW searches for rare transients, weak continuous signals or
a stochastic background, the strain amplitude time series $h(t)$ of
each detector may contain a GW signal buried in the instrumental
noise. The sensitivity of interferometric detectors varies as the
detector noise increases or decreases. Noise events and GW events can
have similar properties and the challenge of a data-quality
investigation is to discard as many noise events as possible in order
to improve the sensitivity of GW searches.

Many astrophysical sources are expected to emit short duration GW
signals, such as: the inspiral and coalescence of binary neutron stars
and/or black holes~\cite{Thorne:1987af}, core collapse
supernovae~\cite{Ott:2009bw}, pulsar glitches~\cite{Kokkotas:1999bd},
newly formed and rapidly spinning neutron
stars~\cite{Lindblom:1998wf}, accreting neutron stars in low-mass
X-ray binaries~\cite{Bondarescu:2007jw}, soft gamma repeater giant
flares, anomalous X-ray pulsars~\cite{Glampedakis:2006apa} and cosmic
(super)-strings~\cite{Damour:2004kw}. When the GW signal waveform is
well modeled, as in the case of the compact binary coalescence (CBC),
template-based matched filtering techniques are used to search for
GWs~\cite{CBC,S6CBC}. Otherwise, robust methods to detect a ``burst''
of excess energy in the detector network are
used~\cite{bursts,S6bursts}. Burst detections are particularly
susceptible to the presence of transient noise events (or
{\em glitches}). Even matched filtering searches are affected by noise
glitches, especially when the templates are of short duration. For
these reasons we must understand the nature and the source of the
glitches in a detector. However the amplitude distributions of these
noise transients do not follow a Gaussian distribution. For many years
in Virgo, starting with the first ``commissioning'' runs, great
efforts have been made to identify and locate transient noise sources
that couple into the output of the detector. In the best case
scenario, provided that a noise source is understood, it is possible
to mitigate the noise in the detector or its environment. This paper
will provide a few examples of such cases. However, for many
transients, the noise source cannot be eliminated or the cause is only
understood after the end of the data acquisition period. Therefore
there is no choice but to exclude short periods of time surrounding
these noise events. We refer to this as ``vetoing''.

In addition to transient GW signals, continuous gravitational wave
(CW) signals are expected to be produced by rapidly-spinning
non-axisymmetric neutron stars. A CW signal is expected to be
contained in a narrow band $\Delta f$ centered on a frequency $f_0$,
which depends on the emission mechanisms at
work~\cite{CWSIGNALS}. Targeted searches for known
pulsars~\cite{Abbott:2009rfa, Abbott:2008fx, Abadie:2011md} (known
frequency, position and spin-down rate) use matched filtering techniques
and are restricted to a narrow frequency band ($\Delta f \sim
10^{-4}f_0$). The search for CW signals with unknown parameters is
performed over a much larger parameter space~\cite{cw_moriond}
(all-sky, GW frequencies between 20~Hz and 2~kHz, and for
several possible values of the spin-down rate) which reduces the
sensitivity of the search. Broad-band increases of the detector noise
level are the first obstacle for CW searches~\cite{CLEANING}. This
paper, however, focuses on narrow-band frequency disturbances called
spectral lines (or lines). The presence of lines in the detector
frequency spectrum can significantly reduce the sensitivity of CW
searches. The origins of several lines in the Virgo sensitivity curve
(figure~\ref{fig:sensitivity}) are well-known. Some of these lines are
associated with resonances of different detector components, including
the mirrors (``drum modes'') or the suspension wires (``violin
modes''). This family of lines is part of the detector design and
cannot be removed from the data. There are also constant frequency
signals which are injected into the detector for calibration and
control purposes. This paper focuses on a second class of lines which
are more problematic since many of them have no identified origin or they
cannot be mitigated easily without degrading the general performance
of the detector. Furthermore, some of these noise lines are not
stationary; they fluctuate in amplitude and frequency, making their
identification more complicated. These non-stationarities can also be
a source of glitches that affect transient GW searches. It is thus
important to track the noise spectral lines, to monitor their
characteristics (frequency, amplitude and variability) to ensure that they do not
cross the frequency band of a known pulsar. Both LIGO and Virgo have
dedicated data-analysis tools to achieve this task and to help
identify the sources of noise lines~\cite{NOEMI,
  1742-6596-243-1-012010, 0264-9381-22-18-S33, 0264-9381-22-18-S18}.

A stochastic gravitational wave background (SGWB) is expected to be
emitted in the early stages of the universe evolution (by
inflation~\cite{Grishchuk:1974ny}, electroweak phase
transition~\cite{Kosowsky:1992rz} and cosmic
strings~\cite{Caldwell:1991jj}) or produced as a consequence of the
incoherent superposition of many astrophysical sources like core
collapse supernovae~\cite{Ferrari:1998ut},
magnetars~\cite{Regimbau:2005ey} or neutron star
coalescence~\cite{regimbau:2005tv}. 
SGWB searches correlate two detector's strains over a wide frequency
range~\cite{Christensen:1992wi, Allen:1997ad} and are also affected by
noise spectral lines. The SGWB search is also sensitive to large
transients which distort the estimation of the detector frequency
spectrum used to measure the signal. The published SGWB search
involving Virgo~\cite{Abadie:2011fx} made use of the data-quality work
described here for transient searches to reject the most noisy time
periods for the analysis.
  
This paper gives an overview of the data-quality studies carried out
during the three Virgo science runs designated as VSR1, VSR2 and VSR3
which occurred during 2007-2010 (see~table\ref{tab:virgo_runs}). Many
noise sources were identified by our investigations and we describe
the actions taken to mitigate noise or the procedure used to veto
remaining noise events. The paper is organized as follows:
section~\ref{sec:virgo} presents the Virgo detector, its sensitivity
to GWs and the different Virgo data-taking campaigns over the
years. Several detector sub-systems are also briefly described. In
section~\ref{sec:detchar}, a summary of the detector characterization
work is given. Section~\ref{sec:glitch} focuses on transient noise
sources. We present the different methods which have been developed to
identify glitches, we list the noise sources which have been
identified and we explain how they couple with the strain output. This
section also describes the actions to remove glitches either at the
detector level or from the data-analysis, with the definition of data
quality flags. Noise spectral lines, which primarily affect CW and
SGWB searches, are discussed in section~\ref{sec:lines}. Methods for
identifying lines are briefly described and we review the main
families of lines. Finally, section~\ref{sec:analyses} shows how the
Virgo detector characterization  work impacts the transient GW and CW
searches involving Virgo data. We conclude with
section~\ref{sec:conclusion}, where we present ideas for improvement
of the detector characterization tools and procedures for the next
generation of GW detectors.

%% file: virgo.tex
%
\section{The Virgo Detector}\label{sec:virgo}

\begin{figure}
  \center
  \epsfig{file=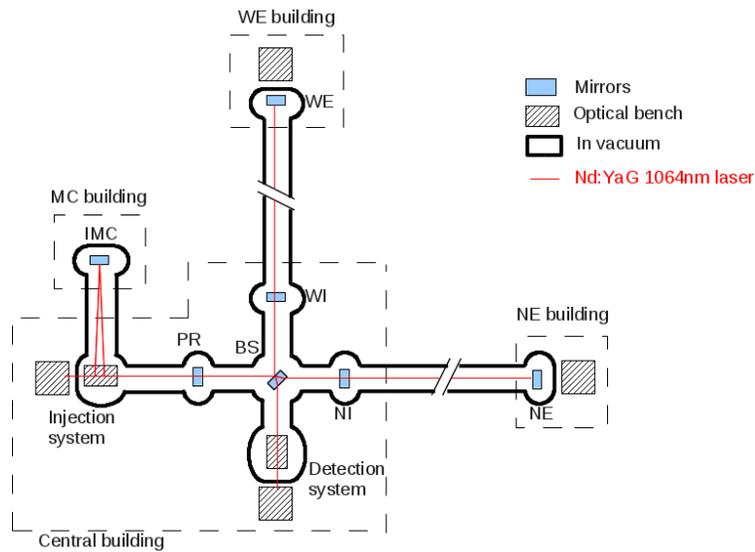,width=10cm,angle=0}
  \caption{The Virgo detector layout showing the main laser path
    through the input mode cleaner (IMC), the power recycling mirror
    (PR), the beam splitter (BS), the western cavity (WI-WE), the
    northern cavity (NI-NE) and the detection system. Most of the
    laser propagation is performed in
    high-vacuum~\cite{The_Virgo_Detector_paper}.}
  \label{fig:virgo_layout}
\end{figure}

Virgo is an interferometric GW detector located near Pisa, Italy,
aiming at directly observing GWs. The optical layout of the detector
is based on a power-recycled Michelson interferometer where each arm
contains a 3-km long Fabry-Perot cavity. The Virgo experiment layout is shown on
figure~\ref{fig:virgo_layout}. An incident GW from a plausible
astrophysical source induces a differential length variation (smaller
than $10^{-18}$~m) between the test masses suspended at both arm
ends. The interferometer is set to operate at a dark fringe and
photo-diodes at the output of the interferometer observe a GW signal as
a fluctuation in the intensity of the light. In the following we will
often refer to the dark fringe (DF) as the uncalibrated GW detection
channel. The calibrated GW strain amplitude, $h(t)$, is reconstructed
taking into account the frequency-dependent transfer functions of the
instrument~\cite{1742-6596-228-1-012015, 0264-9381-28-2-025005} which
are applied to the DF signal. The ability to detect GWs relies on the
stability of the detector, and much attention is given to critical
systems of the instrument: the mirrors, laser and feedback controls.

The isolation of the test masses from seismic activity is crucial in 
order to ensure good sensitivity, especially at low frequencies. In Virgo,
sophisticated super-attenuators (SA)~\cite{SUPERATTENUATOR} have been
installed to decouple mirror motion from seismic fluctuations. A SA
consists of an eight meter chain of five mechanical pendula with a
connection to the ground by three elastic legs, playing the role of an
inverse pendulum. The bottom part of the suspension, called the
payload, is composed of mechanical elements that suspend the mirror
and control its motions. This payload is suspended from the last stage
of the SA. The SA allows for good sensitivity down to 10 Hz.

The main laser (a 1064~nm Nd:YaG laser~\cite{INJECTION}) is a critical
component of the interferometer and special attention must be given to
its stability and its operation. The laser frequency and power are
stabilized and the laser position jitter is controlled to limit the
impact of environmental disturbances~\cite{PhysRevA.79.053824}. A
suspended 144~m triangular cavity, called the Input Mode Cleaner
(IMC), is used to remove high-order modes from the light. The laser
propagation takes place inside a high-quality vacuum to limit air
contamination which could induce phase noise. Scattered light
represents a major contribution to the Virgo noise since it can
recombine with the main laser beam. Such an effect is limited by
installing beam dumps and baffles at strategic points inside the
vacuum tanks~\cite{PhysRevD.56.6085}.

The optical cavities are maintained at resonance using the
Pound-Drever-Hall technique~\cite{springerlink:10.1007/BF00702605},
relying on a laser beam phase-modulated at 6.26~MHz. DC and
demodulated signals from different photo-diodes throughout the detector
are used to control the interferometer. The control loops, running at
a sample rate of 10~kHz, are composed of Analog to Digital Converters
(ADC), and a real-time software architecture that is used to
reconstruct the cavities length. The control system also sends
corrections to the mirror actuators (coils in front of magnets)
through Digital to Analog Converters (DAC) to keep the optical
cavities resonant. Special care is taken to keep the electronic noise
at a very low level and to insure a reliable synchronization between
the different control processes involved in the feedback
systems~\cite{Acernese200829, Accadia2011521}.

\begin{figure}
  \center
  \epsfig{file=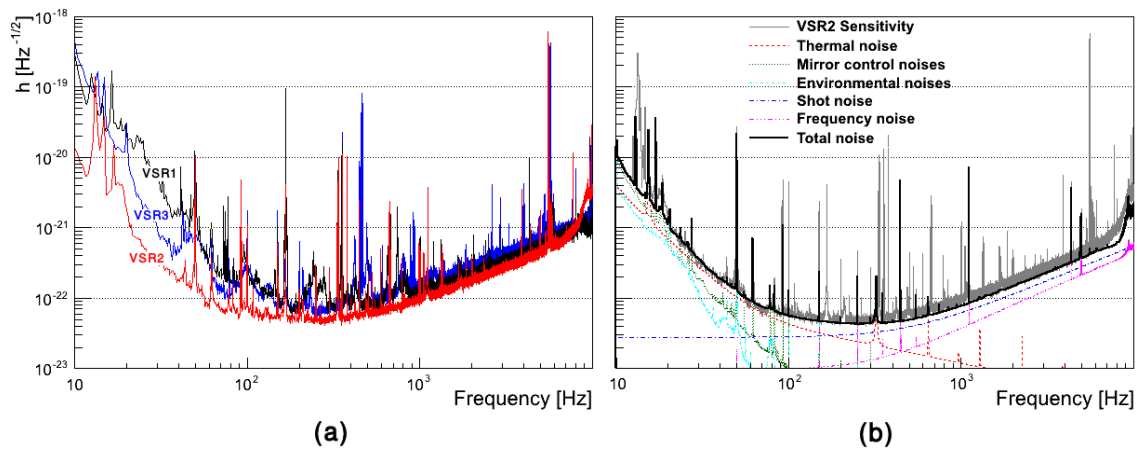,width=15cm,angle=0}
  \caption{(a) Typical sensitivity vs. frequency curves for the first
    three Virgo science runs: VSR1 (2007), VSR2 (2009) and VSR3
    (2010). (b) The measured VSR2 sensitivity curve is compared to
    the predicted noise budget~\cite{NOISEBUDGET}. The agreement between
    the measured and the predicted sensitivity was the best for
    VSR2. For VSR1\&3 the agreement was not as good, especially at low
    frequency.}
  \label{fig:sensitivity}
\end{figure}

Virgo can detect GWs with an amplitude as low as $10^{-21}$ over a
wide frequency band, from tens to thousands of hertz (and below
$10^{-22}$ at a few hundreds of hertz). The sensitivity curves shown
in figure~\ref{fig:sensitivity}(b) are limited by several types of noise
that can be divided into three frequency regions. At low frequencies
(below 100~Hz), the sensitivity is limited by mirror and suspension
thermal noise, mirror control noises, and environmental noises. Mirror
control noise refers to the noise introduced by the feedback systems
used to maintain the interferometer alignment and resonance. This
noise originates from the actuators' electronics and from the control
system's error signals. Environmental noise includes seismic and
acoustic disturbances coupling into the interferometer through
scattered light or input beam jitter, as well as magnetic disturbances
coupling through the mirror magnets. At high frequencies (above 300~Hz)
the sensitivity is primarily limited by the shot noise of the main laser
beam and by laser frequency noise. The frequency noise originates from
the shot noise of the sensor delivering the error signal used in the
laser frequency stabilization. For intermediate frequencies (between
100~Hz and 300~Hz), both thermal noise and shot noise limit the
sensitivity. Noise structures around 165~Hz and 210~Hz are suspected
to originate from scattered light (see
section~\ref{sec:glitch:sources:alignment}).

In addition to achieving a good sensitivity, it is also important to
maintain the detector in operation as long as possible in order to
maximize the live-time (or duty cycle). A lock acquisition
scheme~\cite{Acernese200829, Accadia2011521} was designed to bring
and maintain the Virgo detector to its working point. The Virgo
locking procedure has proved to be very efficient and robust. The lock
can last for several hours or days at a time (see
table~\ref{tab:virgo_runs}). If lock is lost, it can be recovered in a
few minutes. When locked, the detector is manually set in science mode
when a stable state is reached. When in science mode, no external
input or detector tuning is allowed. Science mode ends when decided by
the detector operator (for maintenance or tuning) or whenever an
instability causes loss of lock of the interferometer. The beginning
and the end of a lock segment are considered unsafe in terms of data
quality. Thus, the first 300 seconds after the end of locking
procedure and the 10 seconds of data before the loss of lock are,
\textit{a priori}, rejected and not used for science analysis.

\begin{table}
  \center
  \footnotesize
  \begin{tabular}{|c|c|c|c|}
    \hline
    \textbf{Virgo Science Runs} & \textbf{VSR1} & \textbf{VSR2} & \textbf{VSR3} \\ \hline
    \multirow{2}{*}{Date} &  May 18, 2007 & Jul 07, 2009 & Aug 14, 2010 \\
    &  $\rightarrow$ Oct 01, 2007 & $\rightarrow$ Jan 08, 2010 & $\rightarrow$ Oct 20, 2010 \\ \hline
    Duty cycle (\% of lock time) &  81\% & 80\% & 73\% \\ \hline
    Science time &  108~days & 149~days & 50~days \\ \hline
    Average lock duration &  10~hours & 10~hours & 9~hours \\ \hline
    Max lock duration &  94~hours & 143~hours & 63~hours \\ \hline
    Omega average trigger rate (SNR~$>5$) &  2.1~Hz & 0.6~Hz & 1.8~Hz \\ \hline
  \end{tabular}
  \caption{Virgo runs summary information. Omega~\cite{chatterjiThesis} triggers 
are generated online to estimate the rate of transient noise events.}
  \label{tab:virgo_runs}
\end{table}

The first Virgo science run, VSR1, took place between May and October
2007, in coincidence with the LIGO detectors. The second run, VSR2,
started in July 2009 after a commissioning period devoted to detector
upgrades. These upgrades included: more powerful and less noisy
read-out and control electronics, a new laser amplifier that provided
an increase of the laser power from 17 to 25~W at the input port of
the interferometer, and the installation of a thermal compensation system
(TCS)~\cite{TCS}, to reduce the effects of thermal lensing in the
arms' input mirrors. As a result, the detector sensitivity was much
improved with respect to the previous run, as can be seen on
figure~\ref{fig:sensitivity}(a). VSR2 lasted six months, after which
further upgrades were performed. Higher reflectivity mirrors were
installed to increase the finesse of the Fabry-Perot cavities. As a
test for Advanced Virgo~\cite{ADV}, these mirrors were hung by a new
suspension made of monolithic silica-fibers in order to reduce thermal
noise effects~\cite{0264-9381-27-8-084021}. These detector upgrades
took six months before resuming science with VSR3 from August to
October 2010. The resulting sensitivity in VSR3 was not as good as
expected, however, and was slightly worse than VSR2. It was not
possible to obtain a reliable noise budget in VSR3. It was discovered
that the newly-installed mirrors had a large asymmetry in the
radius-of-curvature and losses. This increased the interferometer's
contrast defect, resulting in higher power in the DF and stronger
couplings to some noise sources. This paper focuses on the detector
characterization work performed during the three first Virgo science
runs. A final run, VSR4, occurred in 2011 for which very few
references will be given in the following. Table~\ref{tab:virgo_runs}
summarizes the performance of the Virgo science runs covered in this
paper.

%% file: detchar.tex
%
\section{Detector characterization}\label{sec:detchar}

The power spectral density shown in figure~\ref{fig:sensitivity} is an
incomplete representation of detector performance as it does not
include transient effects which reduce the sensitivity of GW searches. 
The DF signal can be disturbed by a large variety of noise sources
originating from within the detector or from its environment. The
noise path (or coupling), which connects the noise source to the DF
affects the characteristics of the noise. A long process called
``noise hunting'' consists of tracking down each noise source and
understanding the conversion mechanisms which occur between the source
and the DF. To achieve this task, the Virgo detector is equipped with
hundreds of sensors, including microphones, seismometers,
magnetometers, photo-diodes, current and voltage monitors,
thermometers and cameras. The signals from these auxiliary channels
are used to monitor external disturbances to help determine whether a
candidate event found by a search pipeline was produced by a GW or by
an instrumental artifact. The Virgo noise hunting process can be
summarized as the following:
\begin{enumerate}
\item Identify events (glitches or noise spectral lines), or a family
  of events with similar properties, seen in the DF.
\item Correlate this event with some unusual detector behavior or
  environmental disturbances (human intrusions, earthquakes,
  thunderstorms, etc.).
\item Check the event time against external scheduled events, such as
  the stop/start of infrastructure machineries or changes in the
  interferometer running configuration.
\item An extensive study is performed to tell whether the event
  occurred in time coincidence with an event in one or several
  auxiliary channels. Statistical algorithms are used to quantify the
  correlations between auxiliary channels and the DF, see
  sections~\ref{sec:glitch:flagging} and~\ref{sec:lines:methods} for
  more details.
\item In many cases, the previous studies cannot differentiate whether
  the noise has been identified at its source or somewhere along its
  propagation. Experiments are performed to understand how the noise
  couples into the DF signal. For example, one can artificially inject
  noise in a hardware component and study the response of the
  detector~\cite{GVAJENTE}. Another possibility is to switch off a
  potential noise source to see if the noise disappears. Some examples
  of such actions are given in sections~\ref{sec:glitch:sources}
  and~\ref{sec:lines:sources}.
\item If a noise source is identified, the strategy to remove it
  from the DF is twofold: first we try to eliminate or reduce the
  noise sources; second, we try to reduce the coupling to
  the DF.
\end{enumerate}

An important aspect of detector characterization is reaction
time. When a problem occurs while Virgo is acquiring data, if we can
understand the source of noise quickly, we can make appropriate
modifications to the detector or its environment to mitigate the
noise. To this end, many algorithms are run online which monitor the
detector's data quality. The strain signal is analyzed by various
search pipelines to characterize the type of events that limit the
sensitivity of the searches. Auxiliary signals are monitored in
quasi-real time so as to be able to tell if they are linked to events
found in the GW searches. The loudest glitches and noise spectral
lines are studied, common features are searched for, and cause-effect
relationships are investigated. For VSR2 and VSR3, data was analyzed
shortly after it was collected so the commisioning groups could
mitigate the noise source/coupling as quickly as possible. Depending
on the noise complexity, mitigation actions could last from a few
hours to a few days. Interactions between analysis and commissioning
groups are imperative to make the noise hunting process efficient.

Because many noise sources cannot be clearly understood or mitigated,
they must be identified and tagged in the data. These events will be
vetoed when data are processed by search pipelines with data quality
flags.

For transient searches, data quality investigations consist of defining
lists of time segments of a few seconds long (commonly called DQ flag
segments) where there is a high probability that a glitch is caused by
an instrumental or environmental source. A DQ flag is usually defined
by using an auxiliary signal that indicates that the interferometer
was out of its proper operating condition or that an external
disturbance was present. Any event found during flagged times by the
data analysis pipelines are vetoed~\cite{0264-9381-26-20-204007,
  0264-9381-27-19-194012, 0264-9381-27-19-194010} (see
section~\ref{sec:glitch:flagging}).

For CW searches, data quality investigations consists of tagging,
characterizing and tracking noise spectral lines. Algorithms are used
to establish coincidences between lines in the detector output and
auxiliary channel signals. This information is then used by the search
to reduce the number of false CW candidates (see~\ref{sec:lines}
and~\ref{sec:analyses:cw}).

All of the data quality information is stored in
databases~\cite{VDB,LINEDB}. In addition to reliably archiving data,
the Virgo database may also be used to perform specific queries. DQ
flags and noise lines can be retrieved by analysis pipelines or
through a web interface.

%% file: glitch.tex
%
\section{Transient noise sources}\label{sec:glitch}

\subsection{Investigations}\label{sec:glitch:investigations}

In Virgo, two analysis pipelines are run online,
Omega~\cite{chatterjiThesis} and MBTA~\cite{0264-9381-27-19-194013},
which monitor the data quality for transient GW searches in quasi-real
time. Omega is a burst search algorithm which produces triggers based
on a sine-Gaussian excess power method with frequencies between 48 and
2048~Hz. A discrete Q transform is applied which consists in tiling the
time-frequency plane for a specific quality factor value. For each
tile, it is possible to define a central time, a central frequency, a
duration and a signal-to-noise ratio (SNR) which is simply the ratio of
the total energy content of the tile to the power spectral density of
the detector noise. The Omega algorithm is generic enough to produce
triggers which are a reliable representation of the output of any
transient GW search. Omega is sensitive to typical detector glitches
and provides useful information about the glitch properties. Omega
triggers are often the starting point for glitch investigation, and
special attention is given to high-SNR events. The noise coupling
associated with loud events is expected to be more obvious, and
therefore easier to understand. Moreover, mitigating or vetoing loud
noise events should also remove quieter glitches which are due to the
same noise source. The MBTA pipeline was specifically designed to
detect GWs associated with the coalescence of compact binary
objects. An inspiral waveform template bank is used to match-filter
the data. The intercorrelation between the data and the template,
weighted with the inverse of the noise power spectral density, defines
the event SNR. The glitches detected by the MBTA pipeline are not as
generic as the ones produced by Omega since these glitches mimic the
specific properties of a CBC signal. However, MBTA triggers give a
reasonable sample of the type of glitches that may affect CBC searches.

A glitch detected by Omega or MBTA often results from a sudden
environmental perturbation that then propagates through the detector,
reaches one of the Virgo sub-systems sensitive to this kind of
perturbation, and then couples to the DF signal. For example, an
acoustic disturbance can be converted into mechanical vibrations which
can, in turn, affect optical elements or disturb the main laser
propagation. Auxiliary channels are constantly monitored and analyses
are performed to establish the correlations between glitches in the
auxiliary channels and triggers produced by Omega or MBTA
(glitch-to-glitch identification). In this way, the most relevant
channels are identified, the noise path may be reconstructed, and the
noise sources identified.

Some glitches detected by Omega or MBTA can result from a spectral
line in the DF which becomes non-stationary in amplitude or in
frequency because of fluctuations in the coupling to the noise source
(for instance, alignment fluctuations). This effect can be
particularly harmful for searches using data whitening procedures
(normalization by the detector frequency spectrum) since they amplify
slight amplitude variations of spectral lines. For this type of noise,
a glitch-to-glitch coincidence with auxiliary signals does not
normally identify the coupling and allow us to construct a DQ
flag. However, the frequency of the line can help identify the
noise source and hence the coupling.

\begin{figure}
  \center
  \epsfig{file=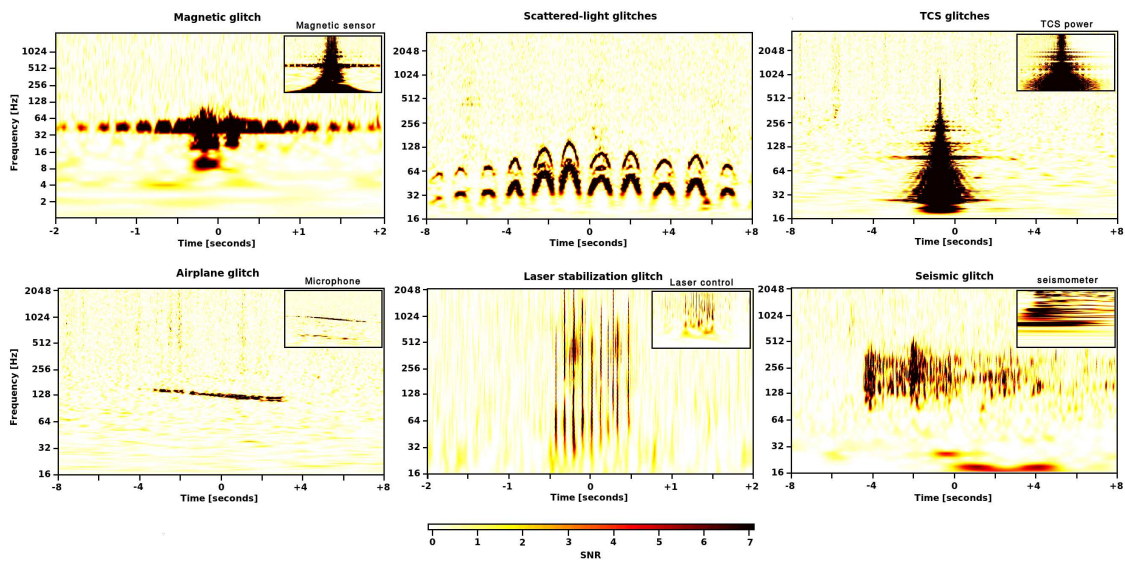,width=15cm,angle=0}
  \caption{Omega time-frequency maps of six examples of glitches seen in the
    DF channel. Glitch families are identifiable by their unique
    time-frequency morphology. When identified, the glitch in the
    auxiliary channel is shown in the inset plot. The first plot shows
    a 50~Hz power-line glitch also detected by the magnetometers. The
    second map shows a series of glitches caused by scattered light
    induced by seismic activity. The third glitch is caused by a TCS
    instability. The fourth plot presents an airplane event with a
    clear Doppler effect. The fifth event is due to a glitch in the
    laser stabilization loop. The last glitch with an undefined shape
    is due to a seismic event up-converted to higher frequencies.}
  \label{fig:omegascan}
\end{figure}

One additional functionality of Omega is its ability to scan a large
number of channels and plot the excess energy as a function of time
and frequency~\cite{chatterjiThesis}. This represents a powerful tool
to identify families of glitches based on the common patterns of the
time-frequency map. Since this process is computationally expensive,
it is typically performed only for the strongest glitches or for a
particular class under investigation. When establishing the
coincidences between channels, it is then possible to reconstruct the
noise path for a given family. Figure~\ref{fig:omegascan} shows an
example of Omega scans of six well-identified families of
glitches. For five of these families, a glitch seen in an auxiliary
channel allowed us to identify the coupling between the noise and the
DF signal.

Most of the detector characterization tools, like Omega scans, were
designed to study glitches resulting from linear couplings between the
noise source and the DF. Within this framework, a noise source can be
identified only if it produces a glitch somewhere on the noise path
that could be detected in an auxiliary channel. For example, the
scattered-light glitches shown in figure~\ref{fig:omegascan} do not
trigger other auxiliary channels. Understanding a noise source outside
this glitch-to-glitch description is a much harder task. Non-linear
couplings are believed to play a major role in the production of noise in
the detector. Only a few of these non-linear noise processes have been
identified and this requires a deep understanding of the experimental
details of the interferometer. Some non-linear couplings will be
described in section~\ref{sec:glitch:sources}.

\subsection{Glitch sources and couplings}\label{sec:glitch:sources}

\subsubsection{Seismic glitches.}\label{sec:glitch:sources:seismic}
Seismic activity is probably the most pervasive source of noise in
Virgo, affecting the detector in many different ways. Almost every
Virgo sub-system is sensitive to sufficiently large
vibrations. Seismic noise can produce a large variety of glitches
which are very difficult to track. Loud seismic glitches due to
violent shocks or earthquakes are likely to produce noise in the
DF. If this happens, the data recorded during a seismic event is
rejected and so the noise coupling is less relevant. The low frequency
signals collected by the multiple seismometers and accelerometers on
site are used to define DQ flags for large seismic activity. Several
frequency bands are monitored at all times, from 0.25~Hz up to
16~Hz. \textit{A priori}, such low-frequency seismic glitches should
not be an obstacle for the transient GW searches whose frequency band
usually starts above 40~Hz. However, seismic noise is often
up-converted in frequency, for example through scattered light
mechanisms as described in~\cite{0264-9381-27-19-194011}. For example,
the seismic glitch presented in figure~\ref{fig:omegascan} was
detected by the seismometers at about 8~Hz and is seen in the DF
signal at much higher frequency ($\sim 200$~Hz).

Bad weather conditions can increase the seismic activity and cause
significantly deteriorated data quality. In such conditions, the
Omega pipeline shows an excess of triggers at low frequency (typically
below 100 Hz). In the case of very bad weather, the Omega trigger rate
below 100~Hz can increase by a factor 5 to 7. During the winter, VSR2
showed many periods of high seismic noise. Substantial efforts were
devoted to studying the resulting
glitches~\cite{0264-9381-27-19-194011}. One family of scattered-light
glitches was characterized by no visible glitch in the Omega
time-frequency maps of the seismic sensors. This fact indicates that a
non-linear coupling was in action. The time-frequency shape of these
triggers is very well-recognizable (see the second plot of
figure~\ref{fig:omegascan}). It consists of a series of arch-shaped
glitches that can last several seconds. The glitches are caused by
light scattered by e.g. the tower walls or the suspended baffles
moving with the micro-seismic motion of the ground. When the
micro-seismic activity is large, higher harmonics can be seen,
probably due to multiple-bounce optical paths. In such conditions,
several rows of arch-shaped glitches can be seen in time-frequency
maps. This noise is well-modeled and the frequency of the arches is
proportional to the velocity of the scattering object. Tests showed
that the position sensors installed at the top stage of the
suspensions are well suited to measure the velocity. The
scattered-light glitches can be rejected when thresholding on the
measured velocity. When applying the DQ flag created in this way, 8\%
of the science time is lost but 2/3 of the scattered-light glitches
were vetoed. The coupling mechanism for the scattered-light glitches
was understood during VSR2 when it was noticed that most of the
scattered light was re-injected into the beam at the level of the
west-end optical bench. For VSR3 the number of scattered light
glitches decreased because of the lower transmission of the new
end mirrors and the installation of absorbing baffles in the west-end
tube. 

\subsubsection{Acoustic glitches.}\label{sec:glitch:sources:acoustic}
Acoustic disturbances can mechanically affect the Virgo systems and
produce glitches. Acoustically-isolating enclosures have been
installed around each optical bench in order to limit the acoustic
coupling with the environment. However, acoustic pollution can either
be produced inside the enclosure or can get inside through mechanical
vibrations. To monitor acoustic noise, each building is equipped with
several microphones. Most of the time, the acoustic disturbances
originate from mechanical devices located near the interferometer
which can be mitigated. However, acoustic noise can also have an
external origin which cannot be controlled or suppressed. Several
times during the day, airplanes or helicopters fly over Virgo and they
are seen in the DF signal. These glitches can be clearly identified by
the typical Doppler shift at about 100~Hz seen in time-frequency maps
of the detector output (see an example in
figure~\ref{fig:omegascan})~\cite{0264-9381-28-23-235008}.

\subsubsection{Electrical glitches.}\label{sec:glitch:sources:mainpower}
Electrical cables represent a major source of noise coupling since they can
propagate an electrical disturbance throughout the Virgo site. The Virgo
sub-systems are usually designed to be electrically isolated from the
environment. However the 50~Hz mains frequency (European standard) can
couple into the detector and transmit magnetic transients. For example, during VSR2,
it was noticed that a family of glitches was periodically produced
roughly every 15 minutes. This effect was identified as electrical
coupling of an air-conditioning unit switching on and off. During the
winter period of VSR2 a loud glitch was produced every day at 8am due
to the heating system that switched on at the beginning of the day and
drew a significant amount of current. Such glitches can be vetoed by
using magnetic sensors that are sensitive to electrical
transients. Electrical glitches are usually corrected by breaking
the electrical noise path. In the two specific cases here, the
glitches disappeared after upgrades to the detector electronics.

\subsubsection{Main laser glitches.}\label{sec:glitch:sources:mainlaser}
One critical element of the Virgo detector is the main laser injection
system. This system contains many control loops to stabilize the laser
power and frequency. Failures in these control systems caused
various families of glitches. During VSR3, the laser power
stabilization control loop was experiencing saturations due to a
mis-tuned gain. This created strong broadband glitches from tens to
thousands of Hz (see figure~\ref{fig:omegascan}). This was fixed a few
days after the problem  was discovered. In the meantime a specific DQ
flag was built to monitor the control loop channel and to efficiently
exclude the glitches from the data attributed to the control failure.

\subsubsection{Dust glitches.}\label{sec:glitch:sources:dust}
Most of the laser light propagation is done in a high-quality
vacuum. However the beam propagates through air in some parts of the
detector, for instance on the injection and detection benches. Some
disturbances, due to dust crossing the beam, for example, create
glitches which are difficult to veto. The laser light propagation can
also be disturbed by unexpected events like spiders building webs or
bugs flying through the beam. It is possible to limit such pollution
by protecting the laser path with plastic covers. Some of the
remaining glitches can be vetoed by using the photo-diode signals of
the secondary beams which are not sensitive to GW signals. Many DQ
flags were created in this way. A very efficient veto was introduced
in VSR1 which relies on the fact that a real GW event seen in the
in-phase demodulated DF channel should not be visible in the quadrature
channel if the demodulated phase is well-tuned. This PQ
veto~\cite{pq_veto} has been extensively used to eliminate these
potential ``dust events'' in the Virgo data.

\subsubsection{Alignment glitches.}\label{sec:glitch:sources:alignment}
The alignment of the main optical beam is critical in order to
maintain the detector in operation. Sophisticated feedback systems are
required to continuously control the optical component angular degrees
of freedom and to optimize the laser beam
alignment~\cite{Acernese2010131,ALIGNMENT}. In the Virgo sensitivity
curves shown in figure~\ref{fig:sensitivity}(a) several spectral lines
are known to correspond with resonances of some optical mounts of the
detection bench (165, 210, 420, 495 and 840~Hz) and are due to light
scattered by these optical components. In principle they should not be
seen as glitches unless they suddenly vary in amplitude which can
happen when the interoferometer alignment conditions change. This
effect of non-stationary lines is a well-known source of glitches to
which transient GW searches are very sensitive. In Virgo, alignment
glitches represent a quite large fraction of Omega triggers (about
25\%). In the case of bad weather, alignment fluctuations are even
larger. As a result, the fraction of glitches due to alignment reaches
40\% and the amplitude of the glitches increases. Alignment signals
can be used to build DQ flags to suppress these alignment
glitches. For VSR2 and VSR3, large deviations of the mirror angular
positions were flagged. This allowed for the removal of as much as
half of the alignment glitches.

\subsubsection{TCS glitches.}\label{sec:glitch:sources:tcs}
The thermal compensation system~\cite{TCS} was installed in Virgo
between VSR1 and VSR2. A TCS instability can directly influence the DF
signal by producing a thermal or a radiation pressure disturbance at
the mirror level. During VSR2, the TCS laser has been stabilized,
reducing the number of glitches. However, it was necessary to
build a specific DQ flag, using the channel monitoring the TCS power,
to veto the remaining glitches. Figure~\ref{fig:omegascan} shows an
example of a TCS glitch that is vetoed by a DQ flag.

\subsubsection{Saturation glitches.}\label{sec:glitch:sources:saturation}
Very loud glitches can be produced by the saturation of different
Virgo active systems. For instance, every photo-diode must operate
within its nominal range ($\pm 10$~V). Specific DQ flags have been
introduced to reject noise transients whenever a photo-diode voltage
is out-of-range. Similarly, the mirror coil driver currents are
monitored to check for saturations.

\subsubsection{Tilt glitches.}\label{sec:glitch:sources:bob}
In data taken two years before the first science run, we identified
a non-linear coupling between the dark fringe and the laser frequency
noise. Laser frequency noise usually lies well below the shot noise
level at high frequencies (see figure~\ref{fig:sensitivity}(b)). Every
27~s, broadband glitches were visible in the DF signal. This period
corresponds to a mechanical resonance in the lower part of the mirror
suspension. The periodic noise increase was correlated with the
extremal angular tilt of the Fabry-Perot cavity's mirrors. When the
mirrors are badly aligned the coupling of the laser frequency noise
increases. To cure this problem, the mirrors' alignment control loops
have been greatly improved. A veto using the direct measurement of the
laser frequency noise in the DF signal (a line at 1111~Hz was injected
in the laser frequency control system) was created to efficiently
eliminate all of the periods containing this noise~\cite{Bizouard:2007}.

\subsubsection{Piezo glitches.}\label{sec:glitch:sources:piezo}
The Virgo detector has many piezo-electric drivers used to control
various elements of the beam path. At the beginning of VSR1, one of
the four piezos of the beam monitoring system was malfunctioning,
causing the input beam to jitter~\cite{0264-9381-25-18-184003}. This
jitter can couple to interferometer asymmetries and was a source of
glitches in the DF signal. The typical frequency of theses glitches
was around 150~Hz. This problem was discovered during the first month
of the run and the piezo was replaced two months later. A similar
problem occurred during VSR3 at the output mode cleaner. A piezo
voltage was found noisy for several hours. The faulty piezo elements
were fixed but DQ flags, based on the control channels, were defined
in order to completely exclude the glitches from the data recorded
while the piezo was faulty.

\subsubsection{Mirror glitches.}\label{sec:glitch:sources:glue}
During VSR2, some glitches were observed with the distinctive feature of
an abrupt step in the $h(t)$ time series which resulted in a loud
broadband disturbance. Theses glitches were demonstrated to be
associated with an excitation of the internal modes of the west-input
(WI) or west-end (WE)
mirrors (depending on the glitches), identified by their accurately
known frequencies. The glitches were interpreted as a sudden
displacement of the surface of those mirrors, of unknown origin. In
the case of the WI mirror, the glitches appeared after the magnets
glued on the back of the mirror were replaced using a type of glue
that had not been used before for that purpose, suggesting the
possibility of a creeping mechanism in the hygroscopic glue. It was
not possible to firmly confirm this suspicion, and the cause of the WE
mirror noise still lacks a convincing explanation. It was impossible
to safely veto those glitches, due to the lack of independent
auxiliary information.

\subsubsection{Thermo-mechanical glitches.}\label{sec:glitch:sources:thermal}
The external temperature can also be an indirect source of
glitches. The steel vacuum tubes, in which the laser travels, have
poor thermal isolation. Hence, external temperature variations are
very likely to mechanically stress the tube through contraction and
expansion. During VSR4, it was understood that when the
expansion/contraction force exceeds the static friction which holds
the tube on its support, a sudden shock occurs and a mechanical
vibration propagates along the tube. This effect is the strongest
around noon and midnight when the temperature gradient is the
largest. Seismometers have been installed to track the noise
propagation and it was found that the noise source was the tube
between the IMC and the injection tower. The resulting glitches are
produced in the DF signal at about 80 and 160~Hz. A seismometer placed
on the injection tower allowed us to flag these glitches with high
efficiency.

\subsubsection{Radio frequency glitches.}\label{sec:glitch:sources:rf}
High-frequency electromagnetic noise overlapping with the laser
modulation frequency (6.26~MHz) can be picked up by the DF photo-diode
signal before demodulation. It can then enter the detector's sensitive
band after demodulation. High-frequency electromagnetic transients are
generated, for example, by fast switching electronic devices
(i.e. power supplies with a typical switching rate of 100~kHz and
above), and by data flow to/from digital devices (the clock rate of
communication protocols is typically in the MHz range). During VSR2
the 6.26~MHz modulation signal was intermittently polluted by a large
amount of glitches that were also seen in the DF signal. The origin of
this noise has never been identified, mostly because of its
intermittent nature. The noise was suspected to originate from serial
transmission devices. It was possible to build a DQ flag based on the
modulation signal to remove the glitches seen in the DF signal. The
beginning of VSR3 showed a large excess of Omega glitches at high
frequency (at 1~kHz and beyond). This was identified as a result of a
coupling between the modulated DF signal and an electromagnetic field
whose frequency was close to the modulation frequency (see
section~\ref{sec:lines:sources:electromagnetic} for more details).

\subsubsection{Digital glitches.}\label{sec:glitch:sources:digital}
The Virgo interferometer is kept at its working point by various digital
control loops. The control servos dedicated to longitudinal control are
fast control loops running at 10~kHz and any digital problem occurring
in theses systems can directly affect the DF signal. One example is a
set of loud glitches in VSR1 that were due to a loss of
synchronization in the control system. This led to dropped samples
between the global control system (which provides the 10~kHz signals
for the interferometer's longitudinal control) and the Digital Signal
Processing board in charge of filtering the correction  signal before
it is sent to a mirror's coil. Combined with a strong but harmless
5~kHz oscillation that is sometimes present in the control signals,
the dropped samples produced loud glitches which were vetoed offline
by searching for missing samples within the 5~kHz noisy time periods.

\subsection{Data quality flagging}\label{sec:glitch:flagging}

In the previous section, the sources of transient noise, which were
identified during the first three Virgo science runs, were listed. As
the sources were understood and localized, the commissioning team
tried to fix the noise sources when possible. However, it was
necessary to create a dedicated DQ flag to veto the glitches before
the fix was performed or when a repair was impossible.

As explained previously, some DQ flags were created by monitoring a
given set of auxiliary channels indicative of noise perturbations
(seismic, acoustic, etc.). The same procedure is used for many DQ
flags: it consists of computing the frequency spectrum of a given
auxiliary channel and to extract the RMS in a specific frequency band
(band-RMS). If this RMS exceeds a given threshold, the data are
flagged as noisy. Many generic seismic flags are generated online in
such a fashion. About 30 seismic sensors are monitored in different
frequency bands: 0.25-1~Hz for the weather conditions, 1-4~Hz for the
car traffic activity and 4-16~Hz for the human activity. Acoustic and
magnetic disturbances are monitored the same way. This kind of
environmental DQ flag does not necessarily point toward a glitch in
$h(t)$ but corresponds to a weaker statement: ``an environmental
disturbance was present in the vicinity of the detector''.

When the noise path to the DF has been understood, it is possible to
use more specific procedures to create a DQ flag that deals with a
category of glitches and which has a great predictive behavior
(measured by the use-percentage defined in
section~\ref{sec:glitch:flagging:perf}). In other words, when a time
period is flagged the probability to find a glitch in the GW data has
to be high. A good DQ flag has to be selective but also efficient (it
must not miss too many glitches of the same
class). Section~\ref{sec:glitch:sources} gives many examples of DQ
flags created to veto specific glitches. Sometimes these flags rely
on a band-RMS where the parameters have to be carefully tuned. In some
cases a simple threshold on the channel value is enough to give good
flag performance. There are also some examples where DQ flags had
to be specifically tailored for a given family of glitches. In
section~\ref{sec:glitch:sources:seismic} we gave the example of the
scattered-light glitches where the velocity of the scatterer was
used to create the flag. Sometimes it is necessary to combine several
channels. One example of this was the DQ flag created to monitor
glitches produced by the large angular deviations of the mirrors
(section~\ref{sec:glitch:sources:alignment}). Multiple mirror degrees
of freedom had to be combined to produce an effective DQ
flag. Finally, in some cases, it has been necessary to use several
channels in time coincidence to provide a DQ flag with good
selection abilities. For instance the 50~Hz glitches, detailed in
section~\ref{sec:glitch:sources}, are usually seen all over the Virgo
site in the magnetic sensors. Therefore the corresponding DQ flag is
defined as a time coincidence between the band-RMS excesses obtained
from several auxiliary signals.

Another method to perform glitch flagging relies on a statistical
approach and does not require any knowledge about the noise source or
the coupling. In this method, we systematically look for noise excess in
many auxiliary channels and correlate it with glitches in the GW
data. For this purpose, the KleineWelle (KW)
algorithm~\cite{kleinewelle} is used to produce triggers for more than
500 Virgo auxiliary channels with a very low latency. As for Omega,
the KW algorithm searches for a statistically significant excess of
power in the time-frequency plane but it relies on a wavelet transform
instead of a Q transform. Omega is known to better estimate the
trigger parameters like the frequency or the SNR. However, Omega runs
much slower than KW which explains why KW was chosen to perform the
auxiliary data analysis. KW triggers are then used by algorithms such as
use-percentage veto (UPV)~\cite{Isogai:2010zz} or hierarchical Veto
(hVeto)~\cite{0264-9381-28-23-235005} to establish coincidences
between triggers of a given auxiliary channel and GW triggers. When
the number of coincidences is much larger than the expected rate of
random coincidences, the channel is selected as interesting in order
to define a powerful veto. By construction, a KW-based veto does not
result from an understood coupling mechanism. For this reason, this
type of vetoes are considered less reliable. This statistical approach
is not only good in terms of glitch flagging but it can also be a
great tool for the glitch investigation. By identifying the auxiliary
channel that best correlates with the DF, this method can help
understand the origin of glitches. In this case, the KW-based veto was
used to construct a DQ flag using a band-RMS of the channel of interest.

Finally, some DQ flags are defined manually by the scientist on shift
in the Virgo control room or at a later time. These DQ flags often
refer to serious detector malfunctions or disturbances in detector
operation. Thunderstorms or earthquakes are systematically reported
and the corresponding time segments are saved for future
reference. The detector operation logbook~\cite{LOGBOOK} is also
carefully examined and when a Virgo sub-system failure is reported,
a specific DQ flag is created. For example, several DQ flags were
defined based on photo-diode, TCS or data-acquisition malfunctions.

In the following, all flags and vetoes described above, including the 
PQ veto, are called DQ flags. When designing a DQ flag, one should
always keep in mind that the flag must not couple to a real GW event
(i.e. the flag is safe for the GW events), while, at the same time, it
must efficiently eliminate noise transients (the flag has good
performance). Those two important aspects are described in the next
two sections.

\subsubsection{Data quality flag safety.}\label{sec:glitch:flagging:safety}

All vetoes, except the PQ veto~\cite{pq_veto}, are derived from channels
that are assumed  to be independent of the DF (which may contain a GW
signal). By accident, a veto can dismiss a genuine GW signal,
but the probability of such an event must be small and follow the
Poisson probability of coincidence between two random processes. To
test that a veto is safe, fake GW signals are injected into the
interferometer by applying a force on one mirror of one Fabry-Perot
cavity to mimic the path of a GW event (hardware
injections). Different types of signals are injected, but to test the
veto safety, the very loud (SNR $\sim$ 100) GW burst signals were used
(Sine Gaussian waveforms with a frequency between 50~Hz and
1300~Hz). These hardware injections, grouped by 10, are regularly
performed at a rate that varies between once a day and once each three
days, depending on the science run. We count the number $N_{flagged}$
of vetoed hardware injections. This number is compared to the expected
number of hardware injections accidentally vetoed:
\begin{equation}
  N_{flagged}^{exp} = \frac{T_{f}}{T_{tot}} \times N_{GW} \ ,
\end{equation}
where $T_{f}$ is the time rejected by the flag, $T_{tot}$ is the total
science time and $N_{GW}$ is total number of hardware injections
performed during $T_{tot}$. The Poisson probability to have
$N_{flagged}$ or more events when $N_{flagged}^{exp}$ are expected is
simply
\begin{equation}
  p (N \ge N_{flagged}) =   \sum_{n=N_{flagged}}^{n= \infty} P(n,N_{flagged}^{exp}) \ ,
\end{equation}
where $P(n,\lambda)$ is the Poisson distribution of mean $\lambda$.
This defines the probability that the veto is safe. Setting a
threshold on this quantity provides an automatic means to determine
which veto is unsafe. Two thresholds were considered: when the
probability is lower than $10^{-5}$, the flag is considered
unsafe. When the probability is below $10^{-3}$, all flagged hardware
injections are manually inspected to determine if this low probability
is due to the fact that a long DQ flag segment has vetoed several
hardware injections belonging to the same series since the hardware
injections are grouped by 10, each separated by 5 seconds. It has been
checked that \textit{a priori} unsafe flags based, for instance, on
channels that are known to contain a fraction of a GW signal have a
probability well below $10^{-5}$. On the other hand, all flags,
\textit{a priori} safe but with a probability between $10^{-5}$ and
$10^{-3}$ were found to be safe, the low probability being due to the
effect explained above.

\subsubsection{Data quality flag performance.}\label{sec:glitch:flagging:perf}

A data quality flag is said to have good performance if it is able to
veto glitches affecting an analysis pipeline without vetoing long
periods without noise transients. DQ flag performance is measured by
considering a set of $N_{t}$ triggers spanning a large frequency band
and the science period $T_{tot}$ of the GW transient searches (the
mean trigger rate is $R=N_{t}/T_{tot}$). Each DQ flag is characterized
by the number $N_{seg}$ of disjoint time segments and the total time
$T_{f}$ rejected by the flag. Three figures of merit, discussed in
details in~\cite{0264-9381-27-16-165023}, are used to measure the flag
performance:
\begin{enumerate}
\item \textbf{The efficiency} ($\epsilon$) measures the percentage of
  triggers vetoed by a DQ flag and is given by $N_{f}/N_{t}$, where
  $N_{f}$ counts the number of flagged triggers.
\item \textbf{The use-percentage} ($U\!P$) gives the fraction of DQ
  segments which are actually used to veto triggers and is given by
  $N_{use}/N_{seg}$, where $N_{use}$ is the number of segments used to
  reject at least one trigger. When this number is close to 1, the
  flagged time period certainly contains a glitch that the DQ flag was
  designed to veto.
\item \textbf{The dead-time} ($D=T_{f}/T_{tot}$) is the percentage of
  science time rejected by a flag.
\end{enumerate}
It is often convenient to compare the efficiency to the dead-time in
order to make sure the flagging is not random. In case of random
flagging, $N_{f}=R\times T_{f}$ and $\epsilon/D=1$. On the contrary,
if the DQ flag is highly selective for glitches,
$\epsilon/D>1$. Finally, a DQ flag has $\epsilon/D<1$ if it tends to
systematically flag periods of time where no triggers can be
found. These figures of merit must be used with care and have
limitations. In particular, they are average numbers and they may be
biased by large variations of trigger rate or by the segment structure
of the DQ flag.

\begin{table}
  \center
  \footnotesize
  \begin{tabular}{|l|l|l|}
    \hline
    \textbf{Category} & \textbf{Definition} & \textbf{Prescription for analyses}\\\hline\hline
    \multirow{2}{*}{\textbf{CAT1}}
    & Flags obvious and severe      & Science data are re-defined when\\
    & malfunctions of the detector. & removing CAT1 segments. \\\hline

    \multirow{3}{*}{\textbf{CAT2}}
    & Flags noisy periods where the coupling & Triggers can be automatically removed \\
    & between the noise source and the DF    & if flagged by a CAT2 veto. \\
    & is well-established.                   & Good performance. \\\hline

    \multirow{3}{*}{\textbf{CAT3}}
    & Flags noisy periods where the coupling & CAT3 flags should not be applied \\
    & between the noise source and the DF    & automatically. Triggers flagged by a CAT3 \\
    & is not well-established.               & veto should be followed up carefully. \\\hline
  \end{tabular}
  \caption{Category definition and prescription for transient GW searches
    (see section~\ref{sec:glitch:flagging:perf} for more details).}
  \label{tab:category}
\end{table}

A DQ flag's performance and the level of understanding of its
corresponding noise source determine at which stage of a GW search the
DQ flag should be applied. DQ flags are divided into three categories:
CAT1-3, defined in table~\ref{tab:category}. When a severe malfunction
prevents the detector from working in normal operating conditions, the
corresponding period must be discarded from the GW searches. Such CAT1
DQ flags are used to re-define the science segments on which analysis
pipelines are run. CAT2 DQ flags are characterized by high performance
resulting from a good description of the noise source and its coupling
with the DF. CAT2 flags can be applied with confidence to the output
of transient GW searches. CAT3 are effective at removing transient
noise from the data, but in the presence of a weak physical coupling,
caution is exercised when using these flags. Furthermore, CAT3 vetoes
typically have an overall larger dead-time than CAT2 flags ($\sim
10\%$). For these reasons, transient GW searches are usually performed
in two steps: in the case of CBC searches the search output, with
both CAT2 and CAT3 flags applied, is first considered to make
statements about the significance of GW candidates or, in the absence
of a detection, to derive upper limits on the GW event rate. If no GW
candidate is observed, GW candidates after CAT2 flags have been applied
are considered. This search is less sensitive, since the noise
background is higher, but it allows to ensure that a significant GW
event has not been vetoed accidently because of a large dead-time
veto. For the burst searches, all triggers after CAT2 flags are taken
into account. CAT3 flags are then used when computing their
significance~\cite{S6bursts}. Some additional DQ flags are
uncategorized, because of very low performances or highly uncertain
coupling mechanisms. These flags are only considered during the
follow-up procedure if a GW event is found to be
significant~\cite{0264-9381-25-18-184006}. Further studies of
auxiliary channels at the time of the event may rule out an
astrophysical origin for the event.
\\

The definition of flags for transient glitches is an iterative
process. Most DQ flags are produced online, but from one run to
another, the noise coupling can change as new noise sources appear
and existing noise is mitigated. The next step is to estimate the
performance of the DQ flags used by each data analysis search in order
to veto transient features and to determine what noise sources remain
after the application of DQ flags. This will be discussed in
section~\ref{sec:analyses:transient}

%% file: lines.tex
%

\section{Noise spectral lines}\label{sec:lines}

\subsection{Investigations}\label{sec:lines:methods}
To describe a spectral line, it is common to use the frequency ($f_0$)
and amplitude ($A$) of the peak maximum. In addition, the line width
($W$) is the peak width at half maximum, the persistence ($P$) is the
fraction of time the line is visible and the critical ratio ($CR$)
is the difference between the peak amplitude and the mean value of the
spectrum, divided by the spectrum standard deviation. Another commonly
used parameter is the line energy, which can be defined as the
integrated power spectrum of the line over its width, averaged over a
given time interval (see \cite{NOEMI} for more details). As stated in
section~\ref{sec:introduction}, we will focus on lines
which, unlike those associated to the interferometer internal modes or
intentionally added for calibration and control purposes, do not have
a well-known origin. Non-stationary lines, with $f_0$ and/or $A$
varying with time, are particularly troublesome since they are likely to
produce transient events or cross the frequency bands of interest
for CW searches.

\begin{figure}
  \center
  \epsfig{file=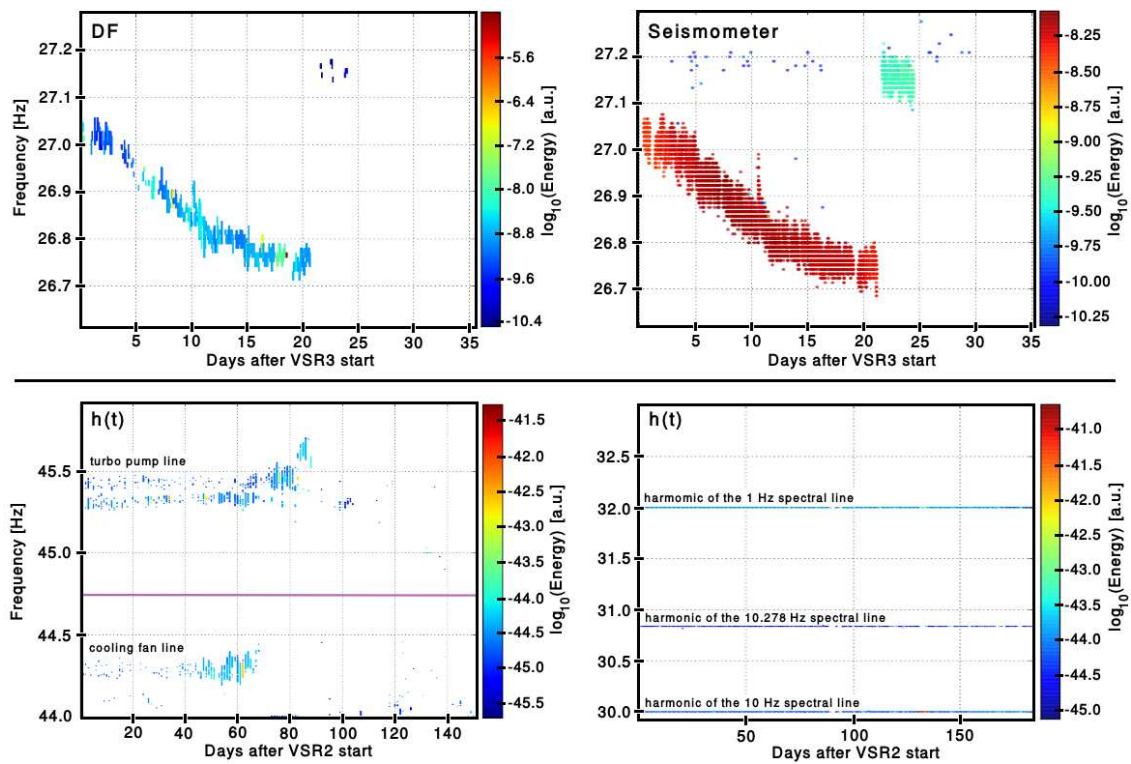,width=15cm,angle=0}
  \caption{ Examples of noise lines reconstructed with
    NoEMi. The color scale refers to the line energy defined
    as the integrated amplitude of the line. The upper row shows the
    coincidence between a noise line seen in the DF channel and in
    a seismometer during VSR3. The correlation between the two lines
    allowed for the identification of the noise source as an air-conditioning
    fan. The lower-left time-frequency plot shows a noise line at
    $\sim$44.3~Hz, which was associated with electronic board cooling
    fans that induce noise currents in the arm mirrors' correction
    signals. The lines around 45.5~Hz were due to the turbo pump
    cooling fans vibrations propagating to the DF through scattered light
    on the vacuum tank walls. These VSR2 lines disappeared after
    the noise source had been mitigated. The lower-right plot presents
    stationary harmonics of digital noise spectral lines (1, 10
    and 10.278~Hz).}
  \label{fig:noemi_lines}
\end{figure}

The line hunting and mitigation process follows the detector
characterization procedure described in section~\ref{sec:detchar}. In
addition to generic line tools~\cite{1742-6596-243-1-012010,
  0264-9381-22-18-S33, 0264-9381-22-18-S18}, a dedicated algorithm,
NoEMi (Noise Event Miner)~\cite{NOEMI}, has been developed to analyze
Virgo data in quasi-real time. NoEMi is based on the algorithms
implemented for the CW search. On a daily basis, it analyzes the
$h(t)$ channel and a subset of auxiliary channels. NoEMi identifies
the noise lines in auxiliary channels and looks for time and frequency
coincidences between the DF and these auxiliary channels. A line tracker
algorithm reconstructs lines over time, facilitating the follow-up of
non-stationary lines. NoEMi displays the latest results
(time-frequency plots of the peak maps, lists of lines, and
coincidences) on web pages and all the lines are stored in a
database~\cite{LINEDB} which can be accessed offline for further
analysis. A web interface is being developed to provide an easy user
access to the database. In addition, NoEMi raises an alarm if noise
lines are detected at or near the frequency band of the known pulsars
of the CW searches. A detailed description of the NoEMi software can
be found in~\cite{NOEMI} and figure~\ref{fig:noemi_lines} shows
examples of lines detected and followed by NoEMi.

A noise spectral line often results from a mechanical or electronic
device operating in a periodic or continuous working cycle. The
resulting noise can be of seismic, acoustic and/or magnetic nature. At
Virgo, such sources are usually part of the service infrastructure
needed for the interferometer operation. This includes machines for
air cleaning and conditioning of the experimental areas, vacuum pumps,
cooling fluid pumps, small cooling fans for electronic devices,
digital clocks regulating data exchange between electronic devices,
and the mains power supply. Because of non-linearities in the line
generation mechanism or in its coupling to the detector, harmonics
(i.e. integer multiples of a line's frequency), as well as linear
combinations of the frequency of various lines may appear in the
spectrum.

In order to identify the cause of a given line seen in the DF signal it
is important to inventory all frequencies occurring on the Virgo
site. The typical frequency of AC electrical motors is a sub-multiple
of the power line frequency (50~Hz in Europe), from 12.5~Hz (8-pole
engine) to 50~Hz (2-pole). Most engines at the Virgo site are
asynchronous which means that their actual rotation frequency is
slightly less than described above (i.e. about 45~Hz for cooling fans
or about 24~Hz for water pumps). Other mechanical frequencies are also
present. For example, the Virgo air-conditioning machines are 4-pole
engines that drive large fans via belt and pulley systems, the fan
speed is set by the pulleys diameter ratio, typically in the range of
6~Hz to 18~Hz. Higher frequency sources can also be found on the Virgo
site. For example, the Virgo ultra-high vacuum
system~\cite{The_Virgo_Detector_paper} makes use of turbo molecular
pumps which rely on a magnetic levitation system to reduce friction;
these run between 600 and 1000~Hz. All these frequencies change
slightly with time; a few percent variations are observed, resulting
from the mains power frequency fluctuations or changing
temperature. Harmonics are also generated, as a consequence of
non-exact sinusoidal motion due to unavoidable mechanical unbalances.

In the next section we will review the main sources and coupling
mechanisms for spectral lines. Since many aspects overlap with the
transient noise, we refer to section~\ref{sec:glitch:sources} for
complementary details.

\subsection{Spectral line sources and coupling}\label{sec:lines:sources}

\subsubsection{Vibration noise.}\label{sec:lines:sources:vibration}
All the machinery operating frequencies constitute a seismic
background due to the engine vibrations. The on-site seismic sensors
reveal a ``forest'' of spectral lines up to 600~Hz, whose amplitude
roughly decreases as $f^{-2}$, meaning they have a roughly constant
energy content. As explained in
section~\ref{sec:glitch:sources:seismic}, seismic disturbances are
likely to couple to the DF through a variety of mechanisms. It is
often not possible to disentangle all of the spectral lines and to
link them to a specific noise source. However some couplings were
identified and are explained below.

One well-known vibration noise path is located in the injection bench
where the laser beam travels a few meters through optical components
for shaping and alignment purposes before entering the
interferometer. Vibrating optics add angular jitter noise to the
beam. Moreover the Virgo in-air input bench has large quality factor
(20-40) resonant modes around 15-20~Hz and 45~Hz which are associated
with the small rigidity of the supporting legs. These frequencies
happen to exactly match the vibration noises of cooling fans (around
45~Hz) and of some vacuum motor fans (18~Hz); the noise is therefore
amplified. Mitigation was attempted before VSR2 by moving fan-cooled
electronic racks to a separate acoustically-isolated room. Existing
optical mounts were also replaced with more rigid ones. By doing this,
the resonant modes were shifted to higher frequencies where the
vibration noise is weaker. Between VSR1 and VSR2, the accuracy of the
interferometer global alignment~\cite{ALIGNMENT} was improved which
also significantly helped reduce alignment noise due to vibrations.

Scattered light often results from vibrating objects such as lenses,
vacuum link windows or vacuum pipe walls. As discussed in
section~\ref{sec:glitch:sources:seismic}, scattered light can be an
important source of noise. If the vibration is periodic, a spectral
line will be visible in the DF signal. As an illustration, the lower-left
plot of figure~\ref{fig:noemi_lines} shows an example of a noise line
caused by scattered light. The line at 45.5~Hz is associated with the
vibrations of turbo pump cooling fans which propagate to the vacuum
tank walls. This noise was mitigated 90 days after the start of VSR2
by seismically isolating the fans from the vacuum tank. 

During VSR1, another source of scattered light was discovered in the
detection system; a glass window used to isolate the detection vacuum
compartment from the rest of the interferometer was acting as an
efficient transducer of seismic and acoustic noise from the external
environment to the detector. To cure this problem before VSR2 the
window was removed and replaced with a larger aperture pipe with an
associated cryogenic pump. Similarly, during VSR2, some light was
scattered back into the interferometer by the main beam output
window. Improving the quality of the window anti-reflection coating
reduced the noise to a negligible level.

Sometimes the noise path of a spectral line cannot be identified. In
this case NoEMi can provide useful hints e.g. by connecting a noise line to a noise
source based on coincidences with auxiliary channels. For example,
during VSR3, the correlation between the frequency variations of a
line in the DF and a line detected in a seismometer allowed for the
identification of the coupling with the vibration of an
air-conditioning fan (see upper row of
figure~\ref{fig:noemi_lines}). This specific spectral line has been
moved out of the detector's sensitive band by reducing the fan
rotation speed.

\subsubsection{Magnetic noise.}\label{sec:lines:sources:electromagnetic}
Electromagnetic (EM) fields produced by electrical systems are
likely to contribute to the noise spectral lines, especially through
their magnetic component. The noise strength will depend on the
intensity of the field, its frequency and the source distance. Virgo
is mostly sensitive to low frequency EM fields (frequencies less than
a few hundreds of Hz) which couple directly into the detector
bandwidth. At higher frequencies, radio-frequency EM fields are
the main source of noise, since a 6.26~MHz frequency is used to
phase-modulate the Virgo laser beam. Hence, it is important to keep
the frequency region within $\pm$ 10~kHz around the modulation
frequency as free as possible of EM noise.

The main magnetic noise entry path is located at the level of the
actuators used to control the mirror positions. These actuators are
made of coils and magnets which can be disturbed by the presence of a
magnetic field and its gradient. During VSR1 small magnetic sources
(e.g. power supplies and cooling fans used for the mirrors local
control electronics) were located a few meters from the mirrors. The
magnetic field radiated by these small magnetic-dipole-like sources
decays quickly with distance (as $d^{-3}$), and so it was sufficient
to move them away by a few meters to reduce their effect to a
negligible level. After VSR1, to further reduce the magnetic coupling
through the actuators, the magnets were replaced with new and less
intense ones. It was also determined that the mirror recoil mass, made
of aluminum, had an amplification effect on the magnetic field
gradient. For VSR3, new recoil masses, made from a dielectric
material, were installed.

During VSR3, another magnetic noise path was discovered in the fans
used to cool down the electronics which compute the mirror position
correction. The magnetic field radiated by the fan motors induced a
noisy current in the correction signals sent to the actuators which
was converted into mirror displacements. This was solved by increasing
the distance between the fans and the electronics without compromising
the cooling efficiency. The lower-left plot of
figure~\ref{fig:noemi_lines} illustrates this problem by showing a
frequency line around 44.3~Hz which results from the coupling between
the fans and the DF. The line disappeared after 70~days, due to the
noise mitigation actions.

\begin{figure}
  \center
  \epsfig{file=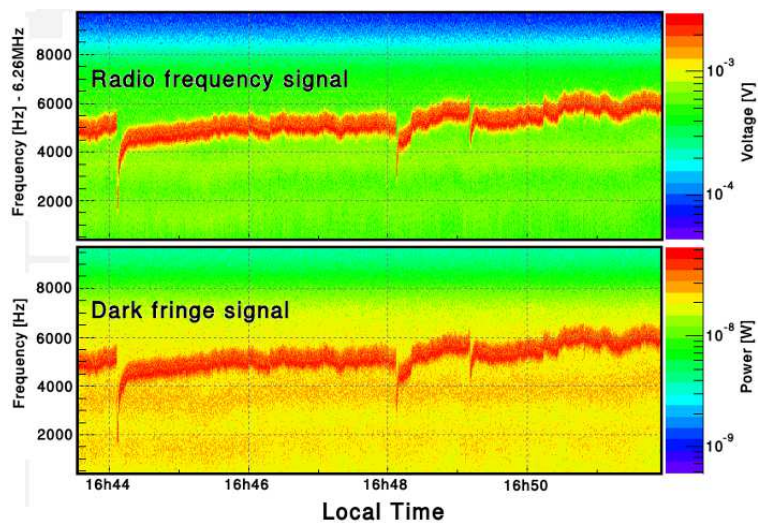,width=10cm,angle=0}
  \caption{Coupling between a local electromagnetic noise source detected by
    an antenna (top) and the DF signal (bottom). The noise frequency
    is very close to the modulation frequency used in Virgo (6.26~MHz)
    which explains why the noise contaminates the DF signal.}
  \label{fig:rf_noise}
\end{figure}

As discussed in section~\ref{sec:glitch:sources:rf}, the
high-frequency EM fields are likely to couple with the modulated DF
signal. During VSR3, an example of a coupling mechanism was
identified: environmental sensor ADCs were using a 300~kHz bit-rate
serial communication protocol, and the $20^{\text{th}}$ harmonic of
this frequency ($\sim 6$~MHz) lies a few kHz from Virgo main
modulation frequency. A radio-frequency antenna showed a 
fluctuating line which was seen in time coincidence with the DF signal
(see figure~\ref{fig:rf_noise}). An unexpected solution consisted of
increasing the temperature of the room hosting the serial link server
by $2^{\circ}$. This slightly changed the clock oscillator rate and
was sufficient to shift the $20^{\text{th}}$ harmonic spectral line
out of the detector's bandwidth.

\subsubsection{Digital noise.}\label{sec:lines:sources:digital}
Another family of spectral lines is composed of very narrow
($W<1$~mHz) and stationary lines at multiples of a few fundamental
frequencies associated with digital systems in the detector. For
instance, several ADC boards used during VSR1 contained a
10~Hz internal clock, which produced a comb of lines spaced by exactly
10~Hz in the frequency domain. These lines were very intense and
covered the whole frequency range of interest for CW searches (between
10~Hz and 2~kHz). A test consisting of switching off the ADCs during
data taking confirmed that they were the source of the disturbance,
although the noise coupling with the DF was not clearly
understood. After the end of VSR1 these ADCs were replaced and almost
all the 10~Hz noise lines disappeared in the subsequent runs. Harmonics of
1~Hz were also observed during VSR1. These lines were concentrated at
frequencies below 100~Hz and tests indicated that the noise source was
probably the same as for the 10~Hz harmonics since this frequency comb
also disappeared nearly completely after th ADC replacement. The
bottom-right plot of figure~\ref{fig:noemi_lines} shows some remaining
harmonics which were still present in VSR2 data despite of the ADC fix.

Another well-known comb of 10.278~Hz harmonics with a digital origin
is present in all Virgo runs. The source of the lines has been
recently identified in digital modules used to control the mirror
coil drivers. There is a strong indication that the coupling mechanism
is of electromagnetic nature.

\subsubsection{Sideband lines.}\label{sec:lines:sources:sidebands}
\begin{figure}
  \center
  \epsfig{file=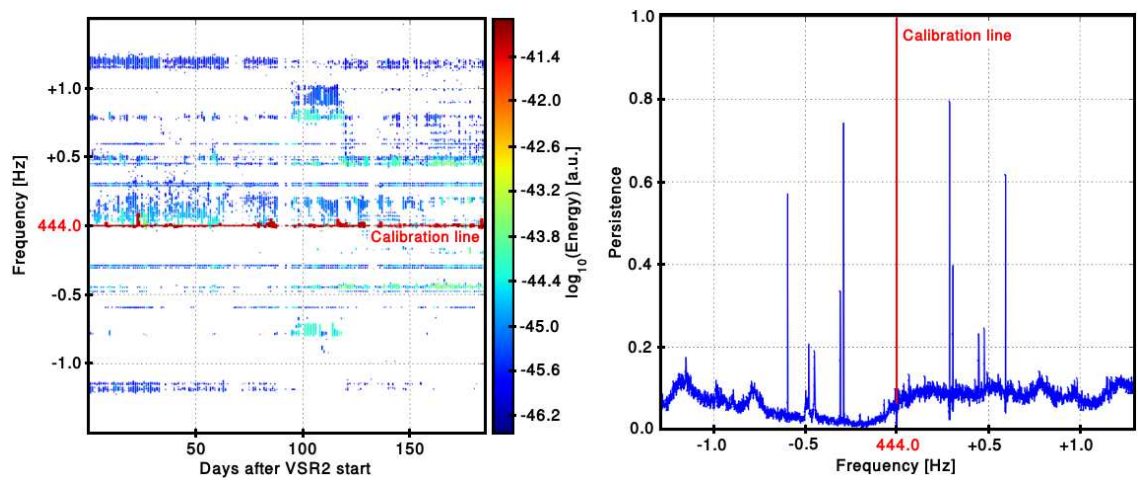,width=15cm,angle=0}
  \caption{Time-frequency (left) and persistence (right) plots of the
    sidebands of the 444~Hz calibration line. The persistence plot has
    been computed over 10 days of VSR2 data, while the time-frequency
    plot covers the full VSR2 run.}
  \label{fig:sidebands}
\end{figure}

The strongest lines in the Virgo spectrum are often surrounded by a
dense forest of sidebands. This effect was identified as a result of a
coupling with the super-attenuator and suspension mechanical
modes. For example, figure~\ref{fig:sidebands} shows the lines
observed on both sides of the 444~Hz injected calibration line and
table~\ref{tab:sidebands} lists the sideband frequencies associated to
each identified mode. The exact mechanism that produces the sidebands
is not known, but is likely due to some non-linearity of the
interferometer.

\begin{table}
  \centering
  \begin{tabular}{|l l|}
    \hline\hline
    $\delta f$ [Hz] & Mode \\
    \hline
    $\pm$0.200 & SA first pendulum mode \\
    $\pm$0.285 & PR payload $\theta_{Z}$ mode \\
    $\pm$0.305 & BS payload $\theta_{Z}$ mode \\
    $\pm$0.450 & SA second longitudinal mode \\
    $\pm$0.595 & SA pendulum mode (last stage) \\
    $\pm$1.200 & BS suspension longitudinal mode \\
    \hline
  \end{tabular}
  \caption{List of identified sidebands associated to the 444~Hz
    calibration line for VSR2 run. $\theta_{Z}$ refers to the angular
    oscillation mode with respect to the beam axis (see also
    section~\ref{sec:virgo} for acronym definitions).}
  \label{tab:sidebands}
\end{table}

%% file: analyses.tex
%
\section{Impact on searches}\label{sec:analyses}

\subsection{Transient GW searches}\label{sec:analyses:transient}

The LIGO-Virgo data are analyzed by multiple search pipelines. This is
motivated by the wide range of GW transient signals expected to be
detected by ground-based interferometers. The use of DQ flags and
their ability to suppress glitches depends on the GW search features,
such as the frequency bandwidth, the use of matched filtering, or
multi-detector coherence tests. In principle, the ability of DQ flags
to remove glitches in the data should be evaluated for each analysis,
and a specific categorization (see section
\ref{sec:glitch:flagging:perf}) should be used. However, to simplify
DQ categorization work, the Virgo detector performance is studied
against only two pipelines: Omega and MBTA. Omega is known to be a good
representation of a burst-type pipeline while MBTA is typical of a
CBC low-mass search.

Although every search is based on a multi-detector analysis, the
performance of a DQ flag is first studied with single detector
triggers. The glitches in Virgo data should be excluded, regardless of
how LIGO and Virgo data are later combined. As a next step,
muti-detector analysis pipelines have ways to estimate the background
affecting a coincident or coherent search (see
section~\ref{sec:analyses:transient:multi}) and we examine the
background triggers coming out of the network analysis pipelines in
order to understand the nature of the harmful glitches. A few
additional DQ flags resulting from this last step were specifically
designed to further reduce the number of loud background triggers.

\subsubsection{Data quality flag performance results.}\label{sec:analyses:transient:perf}

Using the category definition and figures of merit described in
section~\ref{sec:glitch:flagging:perf}, we estimate for each run the
performance of the DQ flags used by GW burst and CBC searches, using
respectively the triggers delivered by single detector Omega and MBTA
online analyses. The list of DQ flags and their category assignment is
then prepared for burst~\cite{S6bursts}, CBC low-mass~\cite{S6CBC}
and CBC high-mass searches~\cite{:2012na}. The prescription can be
different for each analysis. For instance, a DQ flag can be prescribed
as CAT2 for burst and high-mass searches while it is used as CAT3 for
the low-mass search.

\begin{figure}
  \center
  \epsfig{file=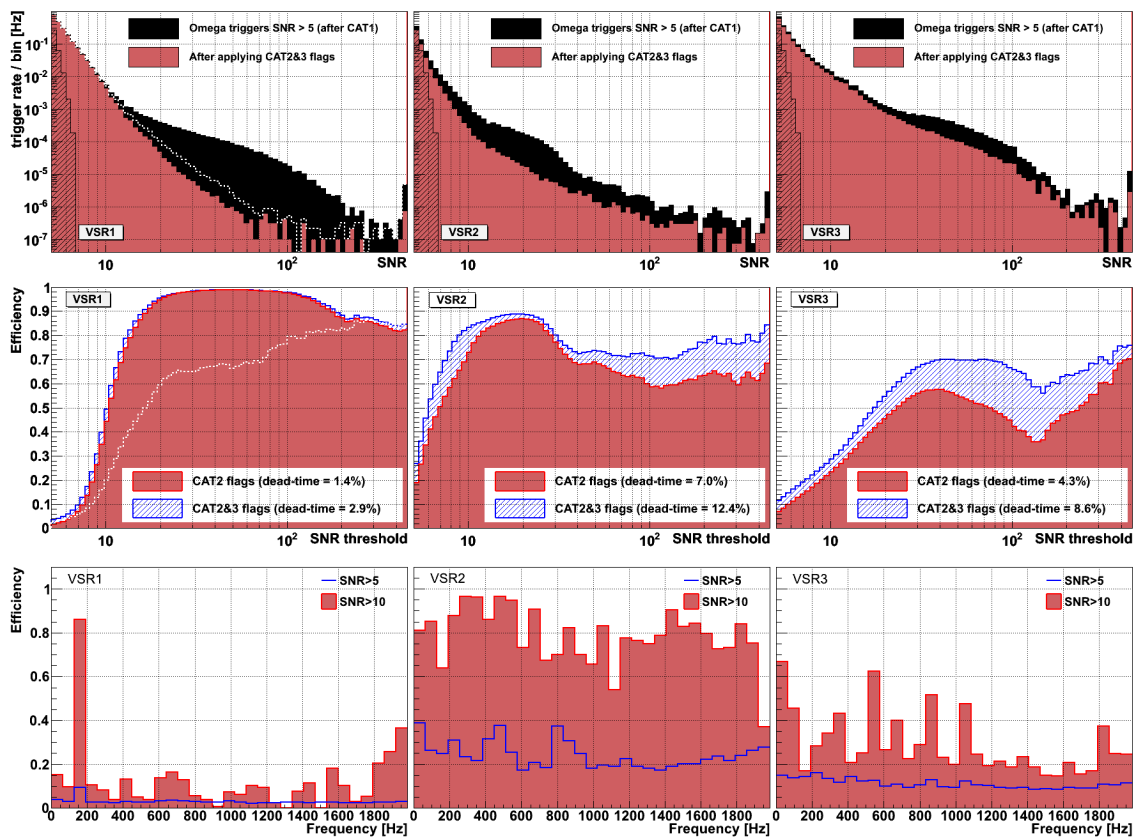,width=15cm,angle=0}
  \caption{Effect of Virgo DQ flags on Omega triggers for each Virgo
    run and after CAT1 flags. In the upper row, the black histogram
    shows the trigger rate in SNR bins while the red distribution
    shows the trigger rate after CAT2\&3 flags. If the VSR1 power
    stabilization flag had been considered as a CAT1 flag, the initial
    trigger distribution would have been given by the dashed white
    histogram. The Omega trigger rate obtained with simulated Gaussian
    noise is represented by the hashed histogram. The middle row
    presents the DQ flag rejection efficiency when considering Omega
    triggers with a SNR above a given threshold. For VSR1, the dashed
    white histogram shows the CAT2\&3 efficiency in the situation
    where the power stabilization flag is considered as a CAT1
    flag. The lower row presents the DQ flag rejection efficiency in
    bins of the central frequency, as determined by Omega.}
  \label{fig:dq_effects}
\end{figure}

Figure~\ref{fig:dq_effects} illustrates the DQ flag performance for
Omega triggers. These results show significant differences between the
Virgo science runs. The first run, VSR1, is characterized by a low
number of DQ flags (about 20) but they are able to reject a large
fraction of loud events. In fact, most of the rejection is obtained by
only one flag and the corresponding noise excess occurred in a single
night of VSR1 when the laser power stabilization failed due to a blown
fuse. The DQ flag was categorized as CAT2 even though the science time
should have been re-defined by removing this noisy period from the
start (CAT1). If one takes into account this correction, the sample of
triggers to be considered is shown by the dashed white histogram in
figure~\ref{fig:dq_effects}. With this consideration, the trigger
rejection is limited and mostly effective for high-SNR
events. Moreover, the initial trigger rate of VSR1 is very high
(2.1~Hz for Omega) and, after applying the DQ flags, remains quite
large (10 times larger than in VSR2 for SNR$>$10).

VSR2 started with a greatly improved knowledge of the detector and of
its response to noise. Furthermore the detector glitch rate decreased
by a factor of 4 with respect to VSR1; this facilitated the noise
investigations. This resulted in a significant increase of the number
of DQ flags (more than a hundred), explaining the larger
dead-time. This also translates into a larger glitch rejection
efficiency: $\epsilon \simeq 70\%$ for SNR$>$8 while it was only
$\epsilon \simeq 10\%$ for VSR1. VSR2 is also characterized by a
rejection efficiency which covers a wider range of SNR. Low-SNR events
are removed with a non-negligible efficiency which is important for
multiple-detector analyses since it is likely that some of the
numerous low-SNR events will combine to produce the most significant
coincidences.

VSR3 data quality was not as good as VSR2 mostly because of the
contrast loss issue explained in
section~\ref{sec:virgo}. Consequently, the detector had to be set on a
new working point which increased scattered-light effects. This
created a significant number of glitches seen by Omega at frequencies
above 500~Hz for which it was not possible to design a DQ flag,
explaining the degraded overall performance of VSR3 DQ flags. As
explained in section~\ref{sec:glitch:flagging:perf}, if the ratio
between the efficiency and the dead-time is larger than one, then DQ
flags target glitches with good accuracy. These numbers can be derived
from the second row of plots in figure~\ref{fig:dq_effects}. For
example, if we consider triggers with SNR$>$10, CAT2\&3 flags give
$\epsilon/D$=6.9, 6.8 and 3.4 for VSR1, 2 and 3 respectively.

Glitch families are often characterized by a given frequency which can
be measured by Omega. It is therefore possible to sort glitch families
and to study the ability of a DQ flag to eliminate them. The frequency
plots shown in figure~\ref{fig:dq_effects} (lower-row) present good flagging
efficiencies in specific frequency bins. For example, the piezo
glitches of VSR1 detailed in section~\ref{sec:glitch:sources:piezo}
are visible at a frequency of $\sim$~140~Hz and are efficiently
rejected. In the VSR2 plot, the efficiency histogram for triggers with
SNR$>$5 exhibits higher efficiency values for frequencies of 210~Hz,
420~Hz, 495~Hz and 840~Hz which correspond to alignment glitches
described in section~\ref{sec:glitch:sources:alignment}. Finally, the
efficiency peaks visible on the VSR3 plot between 500~Hz and 1100~Hz
are mainly explained by the good performance of the laser power
stabilization flag defined in
section~\ref{sec:glitch:sources:mainlaser}.

\subsubsection{Multi-detector analyses.}\label{sec:analyses:transient:multi}

For each trigger, the search pipeline computes a signal-to-noise ratio
statistic after applying a coincidence (CBC) or coherence (burst) test
to determine if the trigger is present in more than one detector. To
measure the background rate of events in the search due to noise, data
from the detectors in the network is time-shifted (by an amount
greater than the gravitational-wave travel time difference between
observatories) and then re-analyzed. Many different shifts are
performed to obtain an accurate measure of the background rate in the
search. The significance of a candidate GW trigger is characterized by
its false alarm rate (FAR), which is computed by comparing the SNR of
the candidate trigger to the background. An excess of noise events in
a detector can cause the distribution of the background to have a
significant non-Gaussian tail at high SNR, thus reducing the
significance of GW events. It is therefore very important to remove
loud background events by mitigating them in the detector, or excluding
them in the analysis with vetoes. Reducing this non-Gaussian tail in
the background is the primary goal of detector characterization, as it
increases the astrophysical sensitivity of the search.

Using several detectors in coincidence presents many advantages. The
most important one is to reduce the number of background triggers and
hence decrease the FAR of GW signals. Initially, it was believed that
the coincidence between detectors would be sufficient to reduce the
detector noise to its Gaussian component. In fact, it has been
realized that searches are limited by accidental coincidences of
transient glitches. Thus, noise investigations and DQ flags are very
important to improve the sensitivity of the searches. Since VSR1,
Virgo data has been used in coincidence with the three LIGO detectors
offering multiple coincidence schemes, from 2 to 4 detectors. As an
alternative to a basic coincidence between detectors, LIGO and Virgo
data can also be combined coherently~\cite{PhysRevD.68.102001}, taking
into account the individual detector's antenna patterns. This approach
provides an optimal detection efficiency since the network is not
limited by the least sensitive detector (at least when combining more
than two detectors).

For a network analysis, the performance of DQ flags can differ from
what has been obtained with single detector triggers, as presented in
section~\ref{sec:analyses:transient:perf}. To study the effect of
Virgo DQ flags on multi-detector searches, we chose to consider the
coincident CBC low-mass analysis~\cite{S6CBC} and the coherent
all-sky burst search~\cite{S6bursts}. Only a subset of the data used
in the published analyses has been considered to quantify the DQ flags
impact. Moreover, only background triggers will be presented in the
following. Finally, LIGO DQ flags are never applied in the following
studies (except CAT1).

\begin{figure}
  \center
  \epsfig{file=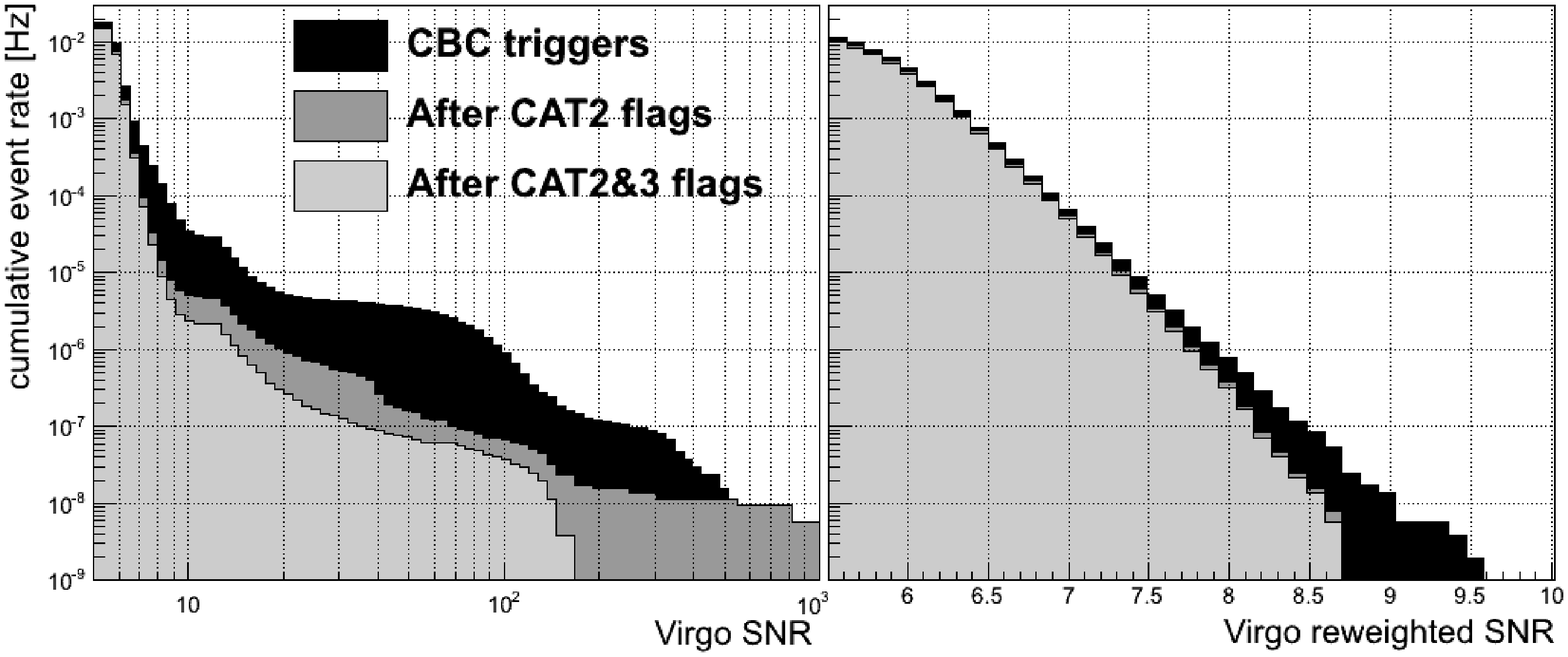,width=15cm,angle=0}\\
  \epsfig{file=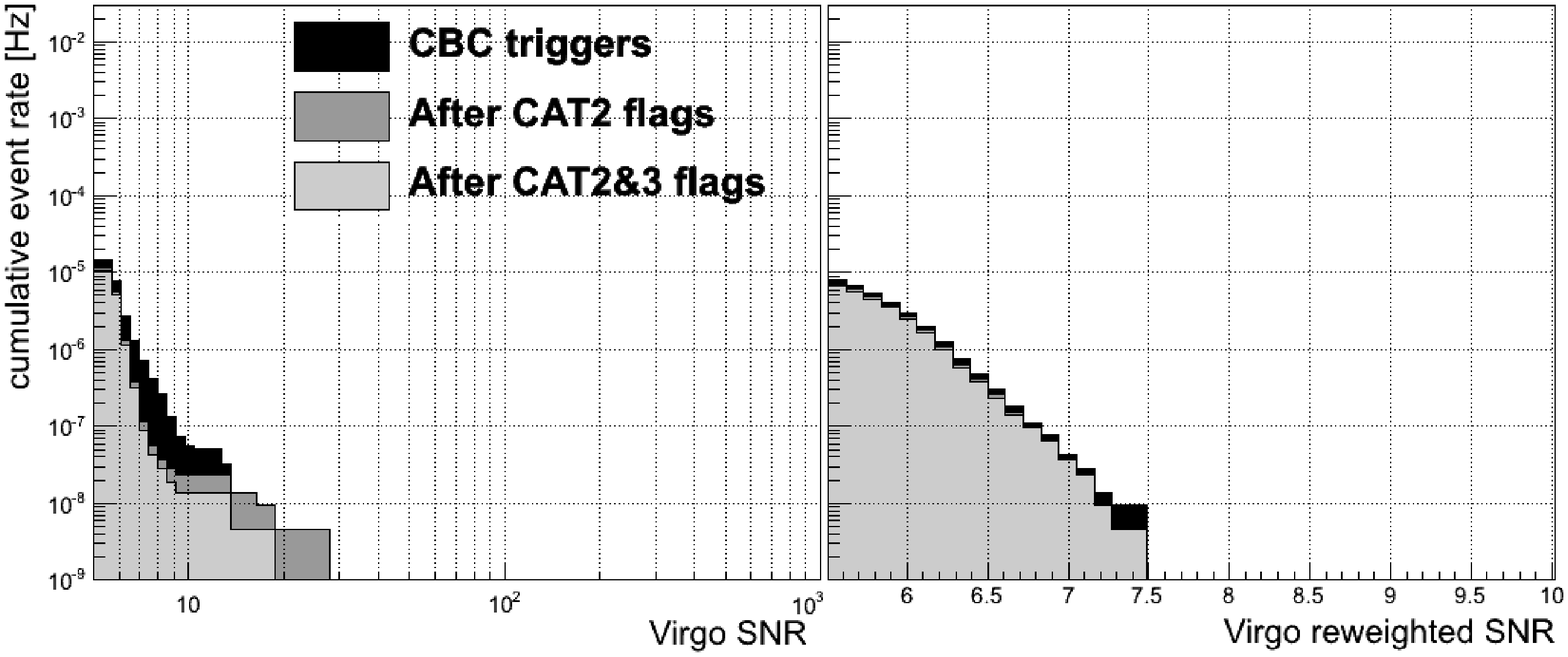,width=15cm,angle=0}
  \caption{Effect of the Virgo DQ flags on low-mass CBC background
    triggers. The full VSR2 data sample was
    considered. Double-coincident (top row) and triple-coincident
    (bottom row) LIGO-Virgo triggers were used. The left column shows
    the trigger rate as a function of the Virgo SNR threshold before
    and after applying CAT2 and CAT3 Virgo DQ flags. On the right
    column, the events are plotted as a function of the Virgo
    reweighted SNR which consists of reweighting the Virgo SNR with
    the reduced-$\chi^2$~\cite{PhysRevD.71.062001} of the event (only
    events with a Virgo reweighted SNR larger than 5.5 are plotted).}
  \label{fig:cbc_perf}
\end{figure}

~

The CBC low-mass analysis makes use of a $\chi^2$ discriminatory
test~\cite{PhysRevD.71.062001} to efficiently reject glitches whose
waveform does not match the expected CBC signal. After having selected
single detector triggers with a SNR larger than 5.5, a preliminary cut
is applied in order to reject events strongly disfavored by the
$\chi^2$ test. For triggers with SNR below~12, an additional cut is
performed based on the behavior of the $\chi^2$ time series near the
trigger time~\cite{Rodriguez:2008kt}. For the remaining triggers, a
reweighted SNR~\cite{S6CBC} is calculated by down-weighting the SNR
progressively with the reduced-$\chi^2$ when reduced-$\chi^2 >
1$. Reweighted SNRs obtained for each detector are summed in
quadrature to form the ranking statistic used in the CBC search. To
evaluate the Virgo contribution to the CBC statistic and the impact of
the DQ flags, both Virgo SNR and Virgo reweighted SNR variables can be
considered.

The upper plots on figure~\ref{fig:cbc_perf} show, for VSR2 data, how
the Virgo DQ flags perform on low-mass CBC triggers which are
coincident in two detectors (Virgo and one of the LIGO detectors). The
combination of CAT2\&3 flags is able to remove background triggers
with an efficiency of 22.6\%. The efficiency increases rapidly with
the SNR measured in Virgo: $\epsilon=92.9\%$ for SNR$>$10 which proves
the ability of Virgo DQ flags to remove the loudest CBC triggers. One
can note that the loudest triggers are removed by CAT3 flags. These
events are found to result from a strong laser disturbance for which a
DQ flag was designed. The performance of this flag is too limited to
be categorized as CAT2. Many events flagged by a Virgo DQ flag are
already disfavored by a high $\chi^2$ value and thus ranked with a low
value of reweighted SNR. Nevertheless, the upper-right plot of
figure~\ref{fig:cbc_perf} shows that Virgo DQ flags have a
non-negligible impact on noise events with large reweighted SNR. For
example, when considering CBC triggers with a reweighted SNR above 8,
approximately 60\% of triggers are removed by Virgo DQ flags. In
general, the SNR of the loudest background event allows us to measure
the sensitivity of a detector, since a GW candidate must be louder
than this to be considered significant. The use of Virgo DQ flags
reduced the reweighted SNR of the loudest event from 9.5 to 8.7,
leading to an astrophysical volume that is 1.3 times larger than the
search without DQ flags. Although here we only consider the
sensitivity of the Virgo detector, and the astrophysical sensitivity
depends on all the detectors in the network, this increase in reach is
a clear indication of the power of data quality and vetoes.

The initial distribution of CBC triggers visible on the upper-left
plot of figure~\ref{fig:cbc_perf} presents some structures which are
understood. First, the SNR distribution shows a steep break at
SNR=12. This effect results from the analysis feature which consists
of applying the $\chi^2$ cuts with a discontinuity at SNR=12. Three
populations of glitches then dominate the SNR distribution. The
loudest events (SNR$>$150) correspond to laser disturbances described
in section~\ref{sec:glitch:sources:mainlaser}. The large bump with
SNR$>$40 results from an excess of TCS glitches (see
section~\ref{sec:glitch:sources:tcs}) which are removed by specific DQ
flags. Events below SNR=12 (which can also be seen as a bump with
Omega triggers in the VSR2 plot of figure~\ref{fig:dq_effects}) are
mostly produced by scattered-light mechanisms described in
section~\ref{sec:glitch:sources:seismic}
and~\ref{sec:glitch:sources:alignment}. The DQ flags based on the
ground motion velocity and alignment signals are able to remove this
population. Moreover, this population of glitches is also
characterized by $\chi^2\sim 1$ (i.e. large reweighted SNR), so these
DQ flags are probably the most important flags to improve the
sensitivity of the CBC search.

The same study has been performed on triple-coincident events (Virgo
and two of the LIGO detectors), as can be seen on the lower row of
figure~\ref{fig:cbc_perf}. The overall performance remains about the
same on triple-coincident CBC events: 30.8\% of efficiency. Requiring
a triple coincidence offers an even more stringent way to suppress the
background than double coincidence, but the reduction concerns all
categories of glitches and thus it does not affect the DQ flags
rejection efficiency. The lower-right plot of
figure~\ref{fig:cbc_perf} shows that the loudest triple background
event of the CBC search is not removed by a Virgo DQ flag.

\begin{figure}
  \center
  \epsfig{file=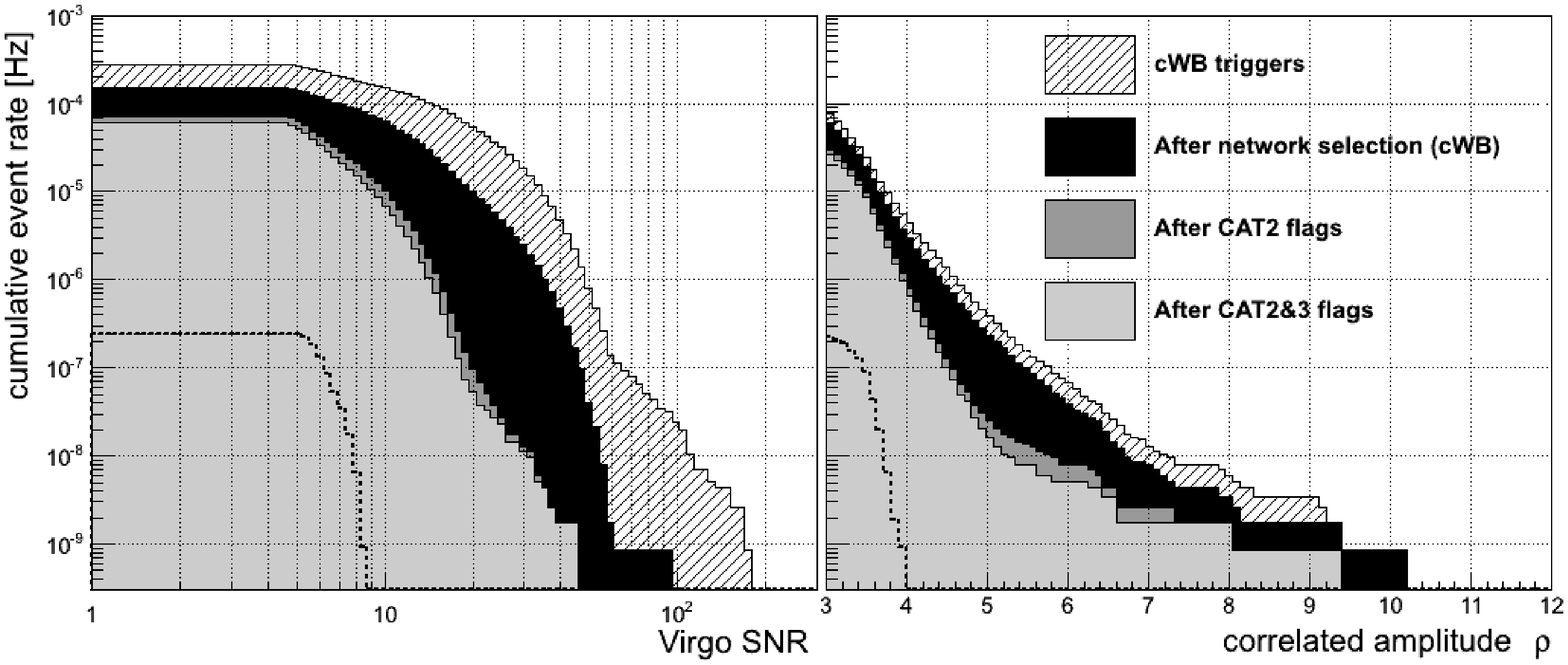,width=15cm,angle=0}\\
  \epsfig{file=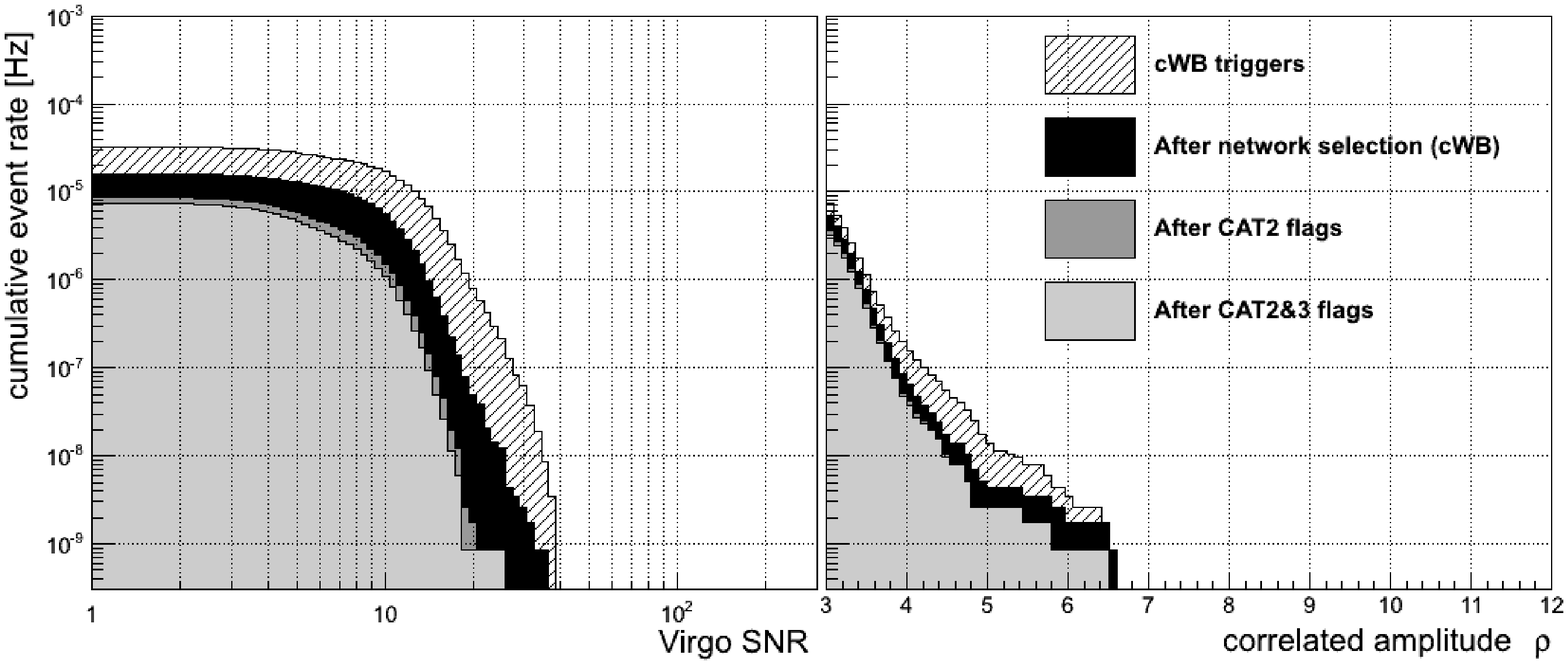,width=15cm,angle=0}
  \caption{Effect of the Virgo DQ flags on burst-type
      (cWB) background triggers. Two months of VSR2 background data
      were considered and coherent LIGO-Virgo event distributions in double
      coincident time (upper row) and triple coincident time (lower
      row) are plotted. On the left, the distribution of Virgo SNR,
      which measures the Virgo contribution to the coherent data
      stream, is plotted and, on the right, events (with $\rho>3$) are
      ranked with the correlated amplitude $\rho$. The hashed histograms show
      the trigger rate produced by cWB. When applying cWB selection
      cuts, it is possible to remove
      loud triggers, as shown by the black histograms. On top of this
      selection, applying CAT2 and CAT3 Virgo DQ flags allows for the
      reduction of the distribution tails even more. The distributions
      obtained with Gaussian noise are superimposed on double
      coincident time plots.}
  \label{fig:burst_perf}
\end{figure}

~

The generic GW burst searches are designed to look for a large variety
of transient signals, spanning the full frequency bandwidth of the
detectors and without a precise model of waveforms. They are therefore
sensitive to a larger number of glitch types than CBC searches and
cannot make use of consistency tests such as the $\chi^2$ test. The
all-sky search~\cite{bursts, S6bursts} has been performed by several
analysis algorithms. Here we use the latest results obtained with the
Coherent Wave-Burst (cWB) pipeline~\cite{0264-9381-25-11-114029} which
combines coherently the detectors' strain amplitudes. In the cWB
search, the network parameters can be derived from a likelihood method
based on a network SNR estimator~\cite{PhysRevD.72.122002} and can be
used to characterize and reject noise transients. Finally, events are
ranked as a function of the correlated amplitude $\rho$, which
measures the degree of correlation between the detectors for an
event. Virgo DQ flags have a significant impact on cWB triggers and
greatly improve the search sensitivity. To study this impact, the cWB
pipeline was run over two months of VSR2 data (2009, November and
December). The winter season of VSR2 was chosen because these data
were the most affected by noise.

In the upper row of figure~\ref{fig:burst_perf}, cWB events obtained
with a two detector network (Virgo and one of the LIGO detector) are
shown. The left plot shows the impact of DQ flags on the distribution
of Virgo SNR which measures the Virgo contribution in the coherent
data stream. Firstly, a collection of selection cuts based on the
likelihood parameters are implemented in the cWB algorithm which
excludes detector glitches incompatible with signals expected from the
detector network. This allows for the suppression of the loudest (and
most obvious) Virgo glitches. Even with this analysis feature,
Virgo DQ flags still efficiently reject part of the remaining
triggers. The overall veto efficiency of CAT2\&3 flags is 60.4\%. For
SNR$>$10, 89.5\% of cWB triggers are rejected by Virgo data quality
flags. The DQ flag rejection efficiency can be derived from
figure~\ref{fig:burst_perf} for any Virgo SNR or $\rho$ threshold when
neglecting the DQ flags dead-time ($\sim 10\%$). Unlike the CBC
analysis, where only a few DQ flags were performing the majority of
the rejection, all Virgo DQ flags contribute to the background
suppression in the cWB search.

As can be seen in the upper-right plot of figure~\ref{fig:burst_perf},
the Virgo DQ flags are less efficient to remove events ranked with a
high $\rho$ mostly because, for these events, the LIGO data strains
preponderantly contribute to the coherent stream. Nonetheless, the
number of high-$\rho$ events is reduced by Virgo DQ flags. For
example, if one fixes the FAR to 1~event per 6~years (rate~$\simeq 5
\times 10^{-9}$~Hz), the cWB network selection cuts allow to lower the
$\rho$ threshold by 10\% and Virgo DQ flags offer an additional 20\%
of reduction. This represents a gain of sensitive volume of about a
factor 2. Such an improvement should be compared with the ideal case
corresponding to Gaussian detector noise (also shown on
figure~\ref{fig:burst_perf}). Data quality work is increasingly
challenging upon approaching this limit. Understanding the glitch
production and coupling mechanisms is much more difficult at lower SNRs.

The same study has been performed on cWB triggers produced with a
three detectors coherent data stream and results are presented in the
lower row of figure~\ref{fig:burst_perf}. As expected, in this
configuration, the search is more sensitive since, for a comparable
FAR, the $\rho$ threshold can be reduced with respect to the two
detector case. For example, with a FAR of 1~event per 6~years, adding
a third detector in the network allows for a 30\% reduction of the
$\rho$ threshold (i.e. the sensitive volume gets twice larger). This
threshold can be further lowered by about 10\% by the use of Virgo DQ
flags (i.e. the sensitive volume gets 30\% larger).

\subsubsection{Online analyses.}\label{sec:analyses:transient:online}

During VSR3, the online data quality monitoring took on a new and
important dimension. Transient GW searches using LIGO-Virgo data were
performed online and alerts were sent to telescopes in order to
observe a possible electromagnetic (EM) counterpart which would
increase the detection confidence of a GW
event~\cite{EM_FOLLOWUP}. Therefore, the data quality information had
to be provided with a very low latency in order to exclude obviously
false GW candidates (noise glitches) which would have otherwise been
sent to telescopes.

For VSR2 an online architecture, based on tools used for the data
acquisition system, was set up to provide DQ flags with a latency of
about 30 seconds. These flags were stored in the LIGO and Virgo
databases. In parallel, the DQ flags were monitored which allowed
scientists in the control room during VSR3 to rapidly check data
quality to make the decision whether or not to send an alert for
prompt EM follow-up.

The main requirements for online DQ flags are: the reliability of the
online production system, the possibility of using the processing
algorithm both online and offline, and the ability to provide a
complete data quality information while at low latencies. During
VSR2 and VSR3, the online DQ production did not encounter major
problems and had a duty cycle similar to the Virgo data acquisition
system (above 99.8\%). The algorithms producing the DQ flags used
generic I/O libraries and thus have also reprocessed missing
segments. Finally, the most difficult part of the DQ flag production
concerns the confidence of the data quality information provided with
low latency. A software architecture has been created to provide
online DQ monitoring. This allowed for the selection of the most
reliable flags in order to veto events before sending alerts to
telescopes. To improve this architecture and to provide accurate
online DQ flags will be one of the main challenges for Advanced
Virgo~\cite{ADV}.

One strong constraint on the online DQ flags is the daily variation of
the glitch rate and glitch types, depending on e.g. the detector
working point or the weather conditions. Online DQ flags performance
can vary significantly if they are not tuned on the
fly. Automatization of such tuning will be an important step to
provide the required reliability of DQ flags for the Advanced Virgo
online analyses.

\subsubsection{Remaining glitches.}\label{sec:analyses:transient:unknown}

The study of data quality is a challenging task and many families of
glitches have origins which have not been identified. For instance,
many Omega scans performed on the VSR2 data show a recurrent glitch
around 60~Hz that always has the same morphology in the time-frequency
plane. It is very likely that these glitches have a common source of
noise. However, no explanation for these glitches has been found.

The lack of understanding of a noise source and of the coupling to the
DF is, in most cases, due to the fact that no auxiliary channel is
correlated with the DF glitches. There are three possible scenarios
which can result in unknown glitch families:
\begin{enumerate}
\item The detector or the environment is not fully monitored: the
  noise source and the coupling mechanism cannot be detected by any of
  the current sensors. This explains why no auxiliary channel has been
  found to be sensitive to this noise.
\item The sensitive channel is actually operational, but it is also
  sensitive to many other kinds of noise which do not affect the GW
  data. In that case, the effective signal component is swamped by
  uninteresting noise and it is highly unlikely that this channel will
  be identified as useful for glitch flagging.
\item The current flagging procedure mostly relies on a
  glitch-to-glitch method. Only a few examples of DQ flags are defined
  by more advanced approaches (for example the scattered-light
  glitches) resulting from a complete understanding of the noise
  path. In the future, it may be necessary to explore more non-linear
  coupling hypotheses (see section~\ref{sec:conclusion} for further
  discussions).
\end{enumerate}

\begin{figure}
  \center
  \epsfig{file=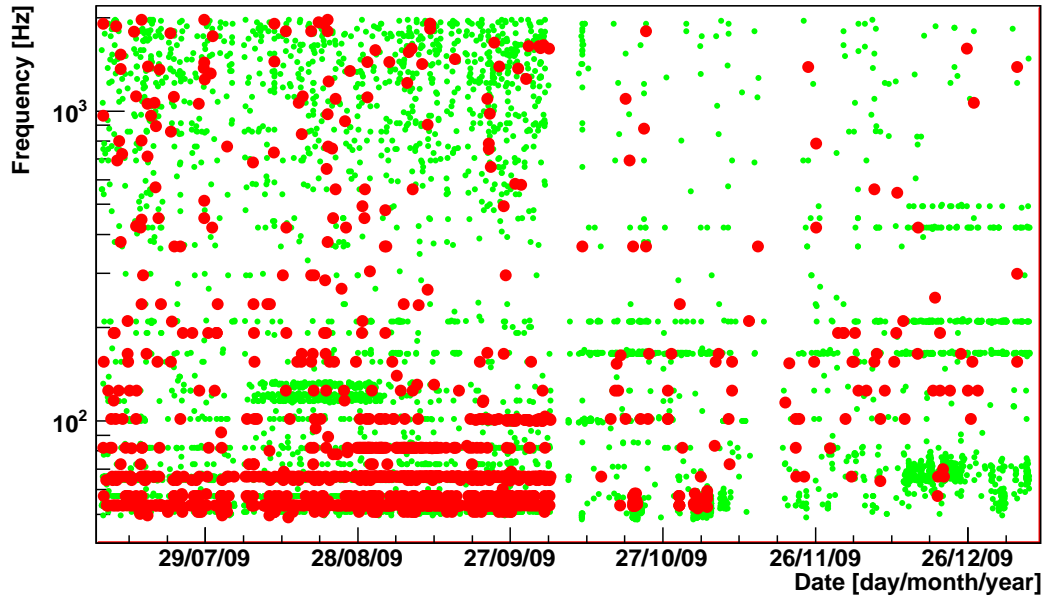,width=15cm,angle=0}
  \caption {
    Time-frequency distribution of the remaining VSR2 Omega triggers
    (from 48 Hz to 2048 Hz) with SNR$>$10 after having applied the
    CAT2\&3 DQ flags (green dots). Triggers with SNR$>$20 are
    represented with a red full circle.}
  \label{fig:remainingtriggers}
\end{figure}

As can be seen on figure~\ref{fig:remainingtriggers}, most of the
remaining glitches do not seem to be associated with a given
permanent noise source that could have been associated with a
specific frequency band. After applying the DQ flags, the low
frequency region remains the most contaminated: triggers with a
frequency below 300~Hz represent 89\% of the remaining triggers. The
``60 Hz glitches'' mentioned above represent about 12\% of the remaining
low frequency glitches. Figure~\ref{fig:remainingtriggers} also
displays a sudden drop of the trigger rate at mid-run. On October
$5^{th}$ 2009, a short commissioning break occurred during which
several actions were performed (dust cover installation, laser and TCS
maintenance) and the exact reason for the glitch rate reduction has
never been well-established.

\subsection{Continuous wave searches}\label{sec:analyses:cw}

\subsubsection{Targeted searches.}\label{sec:analyses:cw:target}

Given the sensitivity of the first generation of interferometers, only
a few known pulsars are astrophysically relevant for close 
examination~\cite{Abbott:2009rfa, Abadie:2011md}. For these pulsars, even
in the case of a null detection, it is possible to approach and possibly beat the
so-called spin-down limit. To achieve this goal, it is important to
make sure that no noise spectral line crosses the frequency band
of these targeted pulsars. This task was performed by the NoEMi software
described in section~\ref{sec:lines:methods}. In
table~\ref{tab:knownpu}, known pulsars monitored in the last Virgo
science runs are listed.

\begin{table}
  \centering
  \begin{tabular}{c c}
    \hline\hline
    Name & $f_0$ [Hz] \\ [0.5ex]
    \hline
    PSR J0835-4510 (Vela) & 22.38 \\
    PSR J0205+6449        & 30.42 \\
    PSR J1833+1034        & 32.31 \\
    PSR J1747-2809        & 38.36 \\
    PSR J1813-1749	  & 44.73 \\
    PSR J1952+3252        & 50.59 \\
    PSR J1913+1011        & 55.70 \\
    PSR B0531+21 (Crab)   & 59.47 \\
    PSR J1400-6325        & 64.14 \\
    \hline
  \end{tabular}
  \caption{Known pulsars monitored by NoEMi. $f_0$ is the expected
    frequency of the GW signal, equal to twice the spin frequency. The
    pulsar frequency bands ($\Delta f \simeq 10^{-4}f_0$) are
    constantly monitored during data taking and an
    alarm is raised if they are contaminated by a noise line. This
    happened during VSR2 and VSR4 runs for Vela, and during VSR4 for
    PSR J1952+3252.}
  \label{tab:knownpu}
\end{table}
 
\begin{figure}
  \center
  \epsfig{file=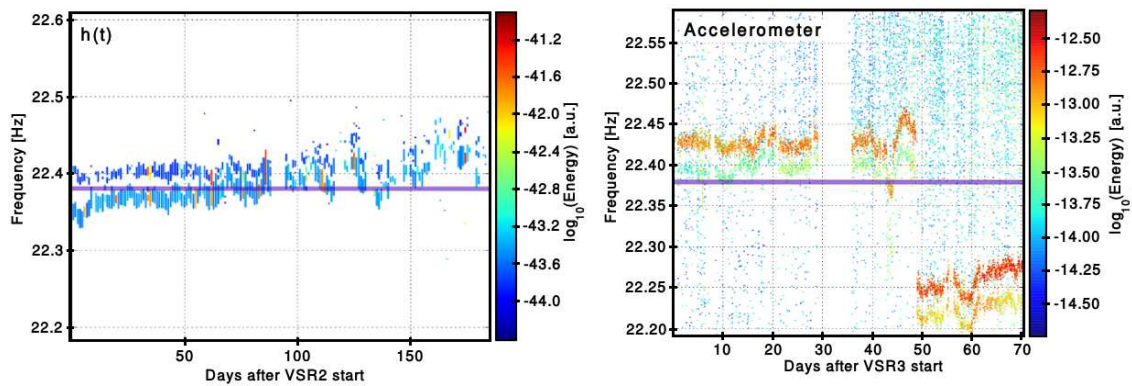,width=15cm,angle=0}
  \caption {
    NoEMi plots showing the Vela frequency band crossed by noise spectral
    lines. The left plot shows the evolution of the noise lines during
    VSR2. The plot on the right shows the same noise detected by an
    accelerometer, which helped to identify the source of the
    vibration disturbance caused by a chiller. The rotation frequency
    of the chiller engine has been changed to move the noise line out
    of the Vela frequency band.}
  \label{fig:velakiller}
\end{figure}

During VSR2 a non-stationary noise line affected the sensitivity of
the Virgo detector at the frequency of the Vela pulsar (22.38~Hz) as
shown on the left plot of figure~\ref{fig:velakiller}. The disturbance
caused a loss of sensitivity of about 20\%~\cite{Abadie:2011md}. Running the NoEMi
coincidence analysis on the auxiliary channels led to evidence
that the disturbance was correlated with a line (actually a doublet of
lines), clearly visible in the data of an accelerometer monitoring the
vibrations of the TCS optical benches (see right plot on
figure~\ref{fig:velakiller}). Although a satisfactory description of
the noise coupling mechanism was not achieved, the source of the
disturbance was identified as being two chillers (pumps that circulate
a cooling fluid for the TCS laser) located near the TCS
room. The rotation frequency of the chiller engine was indeed
22.4~Hz. The vibration was probably transmitted to the TCS bench
through the cooling pipes. During VSR3 the noise line was no longer
visible in the DF, although it was still present in the
accelerometer. It is assumed therefore that the line was hidden under the
detector noise, which at the Vela frequency was 2 to 3 times worse
with respect to VSR2. A small but indicative coherence was indeed
found between the DF and the accelerometer data. To remove the
disturbance away from the Vela band a variable frequency drive was
installed during VSR3 to change the rotation frequency of the chiller
engines, as can be seen in the right plot of figure~\ref{fig:velakiller}.

\subsubsection{All-sky searches.}\label{sec:analyses:cw:allsky}

\begin{figure}
  \center
  \epsfig{file=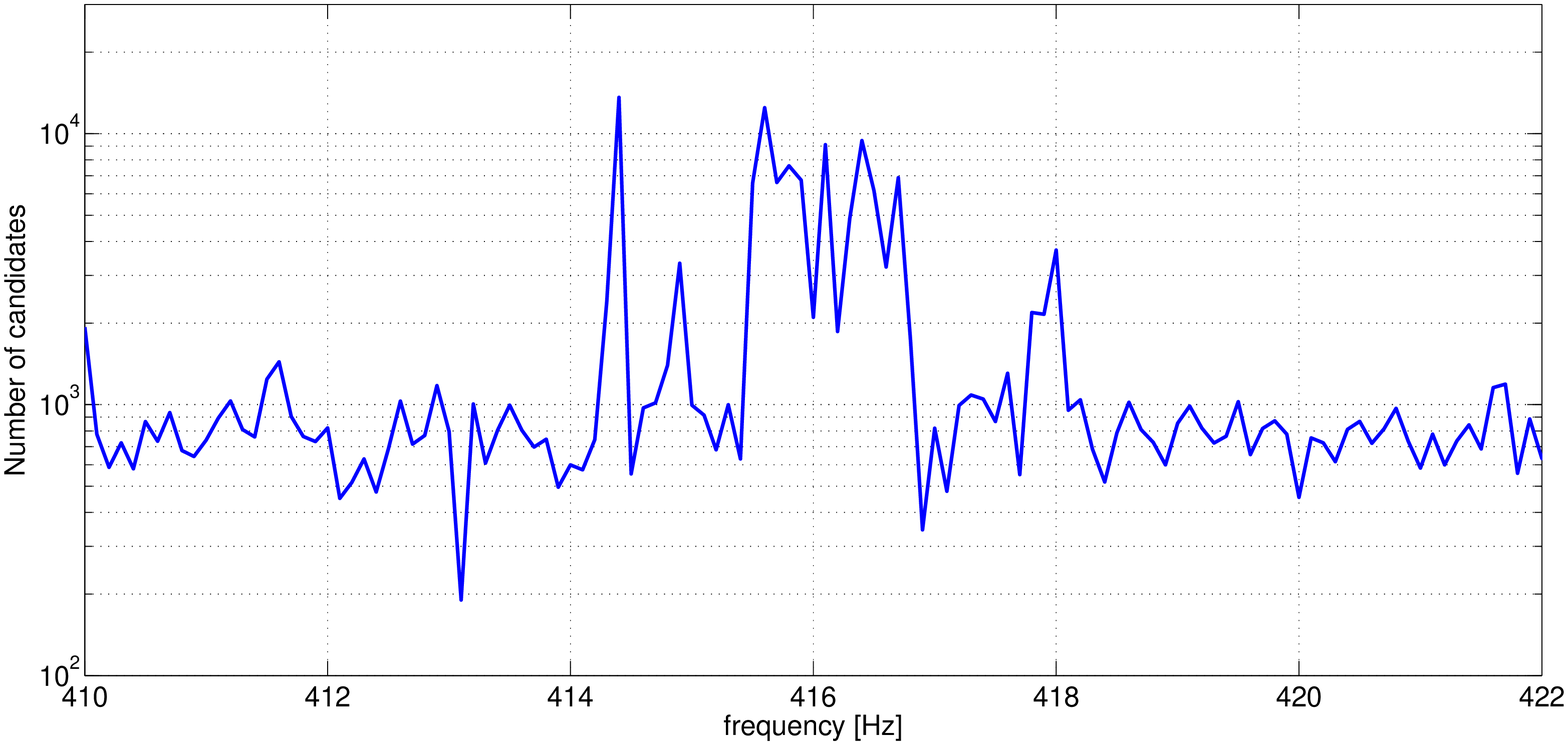,width=12cm,angle=0}\\
  \epsfig{file=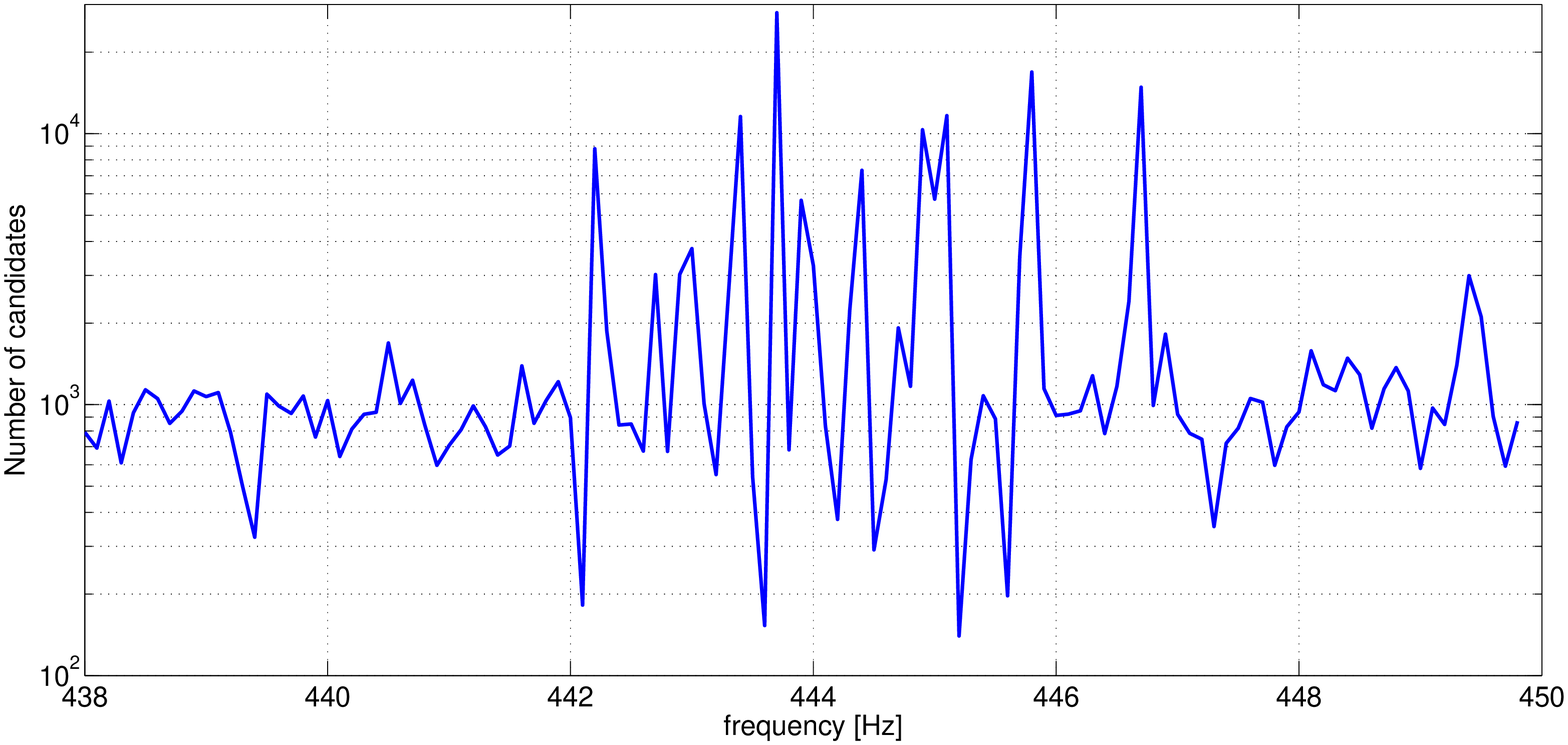,width=12cm,angle=0}
  \caption{All-sky CW candidates found in the 410-422~Hz (top) and
    438-450~Hz (bottom) frequency bands during VSR1
    (bin~width~=~0.1~Hz). The excess of candidates around 416~Hz and
    444~Hz have a known instrumental origin.}
  \label{fig:cand_vsr1}
\end{figure}

All-sky searches produce a list of CW ``candidates'', characterized by a
position in the sky, a signal frequency and one or more frequency
derivatives (spin-down). A follow-up of those candidates is performed
in the next step of the analysis~\cite{0264-9381-22-18-S34,
  0264-9381-22-18-S39, 0264-9381-24-19-S12}. More precisely, candidate
events are selected by thresholding on a quantity characterizing the
candidate significance (using Hough maps built in the source
parameters space). If a noise line is present in the data, it shows up
as a collection of fake candidates. Even a very narrow and constant
frequency  line produces multiple candidates in a frequency band
around it and for various spin-down values. This effect is even larger
in the case of broader lines or a forest of narrow lines, like the
sidebands described in section~\ref{sec:lines:sources:sidebands}. A
sufficiently high threshold on the line significance helps to maintain
a reasonable number of candidates but reduces the sensitivity of the
search. For example, figure~\ref{fig:cand_vsr1} shows the number of CW
candidates selected during VSR1 in the 410-422~Hz and 438-450~Hz
frequency bands. Two excesses of candidates are clearly visible. The
first one, around 444~Hz, is associated with a calibration line and
its sidebands, discussed in
section~\ref{sec:lines:sources:sidebands}. The second excess, around
416~Hz, corresponds to the 10$^{\text{th}}$ harmonic of a 41.618~Hz
noise line and its sidebands. There are strong indications that this
noise line is due to vibrations of the external injection optical bench
producing some beam jitter~\cite{0264-9381-24-19-S07}.

To reduce the number of signal candidates, it is crucial to produce
lists of frequency intervals affected by noise disturbances described
in section~\ref{sec:lines:sources}. One should add to this list all
the lines associated with the intrinsic resonances of the interferometer
as well as the injected lines used for calibration and control. To
achieve this task, all the lines detected by NoEMi are reviewed and
identified one by one. This work is still in
progress. Table~\ref{tab:linesremoved} presents the current status of
the lines identification in VSR3 data. 962 lines have been identified
and about 400 lines still remain to be classified. Once this work is
finished, the frequency bin corresponding to each identified line will
be discarded before running the all-sky CW analysis.

\begin{table}
  \center
  \footnotesize
  \begin{tabular}{|c|c|c|}
    \hline\hline
    & \textbf{Line categories} & \textbf{Number of identified lines} \\ \hline
    \multirow{3}{*}{\textbf{Intrinsic lines}}
    & Violin modes             & 127 \\
    & Mechanical resonances    & 26  \\
    & Calibration and control  & 32  \\ \hline
    \multirow{5}{*}{\textbf{Noise lines}}
    & Power line and harmonics & 40 \\
    & Vibration                & 24 \\
    & Magnetic                 & -- \\
    & Digital                  & 73 \\
    & Sidebands                & 640 \\
    \hline
  \end{tabular}
  \caption{Number of identified lines (over a total of 1390 lines) for
    each category in VSR3 data.}
  \label{tab:linesremoved}
\end{table}

%% file: conclusion.tex
%
\section{Conclusion and perspectives}\label{sec:conclusion}

For several years, the Virgo detector has been operational either in
commissioning mode, for various enhancements and tunings, or in
science mode for four scientific runs in coincidence with the LIGO and
GEO detectors. During this time, investigations have been performed to
improve the overall detector sensitivity and the noise
stationarity. Significant efforts have been made to understand and
mitigate the sources of noise transients and spectral lines that
reduce the sensitivity of GW searches. Virgo noise events present in
the data have been efficiently rejected by defining DQ flags or by
tracking noise spectral lines. Such work has provided good results and has
improved the astrophysical reach of each scientific run.

Since the first scientific run, VSR1, a set of vetoes, mainly based on
DQ flags, have been set up using the monitoring and investigations
made on glitches detected by the online analyses MBTA and Omega. The
DQ flags defined for Virgo data have shown a high rejection efficiency
for noise transients and significantly enhanced the sensitivity of CBC
and bursts multi-detector searches. For instance, during the VSR2 run,
Virgo triggers with SNR$>$10 were rejected with an efficiency of
92.9\% and 89.5\% for CBC low-mass and all-sky bursts searches
respectively. It has been shown that the level of glitch rejection
achieved by this work has allowed a significant reduction in the noise
background of the transient searches. Although a full study of the
astrophysical sensitivity of the network is outside the scope of this
paper, these improvement in data quality can significantly increase
the sensitivity of searches for GWs.

CW search sensitivity can be significantly improved by removing noise
spectral lines contaminating the data. A monitoring tool
(NoEMi~\cite{NOEMI}) has been developed in order to spot and track
noise spectral lines. Whenever possible, lines whose source were
identified during the data taking were mitigated or removed from the
detector's data. All lines are stored in a database which can be
accessed offline to work on their identification. The frequency bins
overlapping with identified lines were eventually vetoed in the CW
all-sky analysis.

Over four scientific runs, the characterization of Virgo data quality
provided for a deep understanding of the Virgo detector and the
properties of its noise. Many noise sources and couplings to the DF
have been fully described. It has been realized that data quality is
an essential feature of the data analysis process, without which it is
impossible to distinguish GW events from the data with sufficient 
confidence. All the tools developed for glitch and line hunting taught
us much, not only about the detector and its various noise coupling
paths, but also how the noise hunting, mitigation and flagging should
be conducted. We also acknowledged the limitations of our current
noise characterization procedures. 

~

The experience gained with the first generation of detectors will
be a great asset when applied to the up-coming Advanced Virgo
experiment~\cite{ADV}, even if noise sources and couplings are
expected to significantly differ from those of Virgo. Before resuming
science in 2015, many projects are expected to be developed in order
to improve the detector characterization methods and to optimize the
use of the data quality information in GW searches. Firstly, DQ flags
used by transient GW searches could be better optimized. For example,
the duration of flagged segments could be adapted to the glitch type
they target. Auxiliary channel monitors also need to be finely tuned
and, for that purpose, investigation campaigns are foreseen to take
place before scientific runs. Finally, the use of DQ flags by search
pipelines could be revisited. Efforts will be needed to develop tools
able to prescribe a DQ flag category specifically tailored for a given
GW search. More ambitiously, all the data quality information could be
combined into a single parameter assigning a probability for an event
to be an instrumental glitch. This parameter could then be folded in
the ranking statistic of every transient GW search. For the line
hunting, NoEMi will be further improved. Some tasks, like the
identification of lines belonging to the same family (e.g. the
sidebands mentioned in section~\ref{sec:lines:sources:sidebands} or
the combs of digital lines described in
section~\ref{sec:lines:sources:digital}), will be automatized. For the
all-sky CW analysis, it is foreseen that the search will be conducted
with a higher frequency resolution. This implies the need to increase
the resolution of the noise line analysis, and therefore the
capability to manage a larger number of noise lines.

New tools for noise investigations are currently being studied. For
example, glitch classifiers and multi-variate analyses represent a
promising improvement for detector
characterization~\cite{0264-9381-24-19-S32,
  0264-9381-28-15-155001,Costa:2011jc}. As stated in
section~\ref{sec:analyses:transient:unknown}, non-linear couplings
will require to be better studied. They are strongly suspected to be a
major glitch production mechanism. Very few tools are available to
systematically track such effects. Nevertheless, many other effects
remain uncovered: slowly time-drifting signals, signal derivative,
signal cancellations, linear combinations of auxiliary signals,
etc. Along the same line of investigation, and as stated in
section~\ref{sec:glitch:investigations}, short time scale
non-stationary lines or couplings are sources of glitches and a tool
will be specifically needed for them.

~

For advanced detectors, online analyses will play a major role. The
improved sensitivity of the detectors will provide access to
many more promising targeted sources among the known pulsar population
for CW searches (of the order of 100)~\cite{MNR:MNR18818}. Realistic
estimations anticipate that $\sim$ 40 binary neutron star coalescence
events should be detectable by the advanced LIGO-Virgo detectors
network per year~\cite{0264-9381-27-17-173001}. Alerts will be sent to
telescopes or satellites for electromagnetic follow-up as soon as
significant GW transient candidates are detected. It is therefore
mandatory to provide the most efficient and reliable online data
quality information as possible. Data quality online architectures
have been tested since VSR2 for both noise line and transient
events. Pulsar frequency bands have been kept under close surveillance
and DQ flags have been produced with a latency of about 30~s. Online
monitoring will be further improved with the addition of new tools for
glitches and lines identification. It will help to provide fast
identification followed by mitigation or veto of noise transients and
spectral lines. Several projects are already in progress in order to
perform detector characterization as reactively and quickly as
possible and to coordinate efficiently the data quality operations.

Many projects and hard work will be needed to improve the detector
characterization and to optimize the use of the data quality
information for Advanced Virgo. A decisive era for GW physics is about
to begin, where reliable and reactive data quality information will
represent a key element to grant due confidence to the first GW event
detection.

%% file: acknowledgments.tex
%
\section*{Acknowledgments}\label{sec:acknowledgments:acknowledgments}

The authors gratefully acknowledge the support of the United States
National Science Foundation for the construction and operation of the
LIGO Laboratory, the Science and Technology Facilities Council of the
United Kingdom, the Max-Planck-Society, and the State of
Niedersachsen/Germany for support of the construction and operation of
the GEO600 detector, and the Italian Istituto Nazionale di Fisica
Nucleare and the French Centre National de la Recherche Scientifique
for the construction and operation of the Virgo detector. The authors
also gratefully acknowledge the support of the research by these
agencies and by the Australian Research Council, 
the International Science Linkages program of the Commonwealth of Australia,
the Council of Scientific and Industrial Research of India, 
the Istituto Nazionale di Fisica Nucleare of Italy, 
the Spanish Ministerio de Econom\'ia y Competitividad,
the Conselleria d'Economia Hisenda i Innovaci\'o of the
Govern de les Illes Balears, the Foundation for Fundamental Research
on Matter supported by the Netherlands Organisation for Scientific Research, 
the Polish Ministry of Science and Higher Education, the FOCUS
Programme of Foundation for Polish Science,
the Royal Society, the Scottish Funding Council, the
Scottish Universities Physics Alliance, The National Aeronautics and
Space Administration, the Carnegie Trust, the Leverhulme Trust, the
David and Lucile Packard Foundation, the Research Corporation, and
the Alfred P. Sloan Foundation.

%% file: paper.bbl
\providecommand{\newblock}{}
\begin{thebibliography}{10}
\expandafter\ifx\csname url\endcsname\relax
  \def\url#1{{\tt #1}}\fi
\expandafter\ifx\csname urlprefix\endcsname\relax\def\urlprefix{URL }\fi
\providecommand{\eprint}[2][]{\url{#2}}

\bibitem{1742-6596-120-3-032010}
Arai K and the TAMA~Collaboration 2008 {\em JPCS\/} {\bf 120} 032010

\bibitem{0034-4885-72-7-076901}
Abbott B {\em et~al.\/} 2009 {\em Rep. Prog. Phys.\/} {\bf 72} 076901

\bibitem{0264-9381-27-8-084003}
Grote H and the LIGO Scientific~Collaboration 2010 {\em Class. Quantum Grav.\/}
  {\bf 27} 084003

\bibitem{The_Virgo_Detector_paper}
Accadia T {\em et~al.\/} (Virgo Collaboration) 2012 {\em JINST\/} {\bf 7}
  P03012

\bibitem{Thorne:1987af}
Thorne K~S 1987 Gravitational radiation {\em Three Hundred Years of
  Gravitation\/} ed Hawking S and Israel W (Cambridge; New York: Cambridge
  University Press) pp 330--458

\bibitem{Ott:2009bw}
Ott C 2009 {\em Class. Quantum Grav.\/} {\bf 26} 204015

\bibitem{Kokkotas:1999bd}
Kokkotas K~D and Schmidt B 1999 {\em Living Rev. Rel.\/} {\bf 2}

\bibitem{Lindblom:1998wf}
Lindblom L, Owen B and Morsink S 1998 {\em Phys. Rev. Lett.\/} {\bf 80}
  4843--4846

\bibitem{Bondarescu:2007jw}
Bondarescu R, Teukolsky S and Wasserman I 2007 {\em Phys. Rev. D\/} {\bf 76}
  064019

\bibitem{Glampedakis:2006apa}
Glampedakis K, Samuelsson L and Andersson N 2006 {\em Monthly Notices of the
  Royal Astronomical Society\/} {\bf 371} L74--L77

\bibitem{Damour:2004kw}
Damour T and Vilenkin A 2005 {\em Phys. Rev. D\/} {\bf 71} 063510

\bibitem{CBC}
Abadie J {\em et~al.\/} (LIGO Scientific Collaboration and Virgo Collaboration)
  2010 {\em Phys. Rev. D\/} {\bf 82} 102001

\bibitem{S6CBC}
Abadie J {\em et~al.\/} (LIGO Scientific Collaboration and Virgo Collaboration)
  2012 {\em To be published Phys. Rev. D\/} (\textit{Preprint}
  \eprint{arXiv:gr-qc/1111.7314})

\bibitem{bursts}
Abadie J {\em et~al.\/} (LIGO Scientific Collaboration and Virgo Collaboration)
  2010 {\em Phys. Rev. D\/} {\bf 81} 102001

\bibitem{S6bursts}
Abadie J {\em et~al.\/} (LIGO Scientific Collaboration and Virgo Collaboration)
  2012 {\em To be published Phys. Rev. D\/} (\textit{Preprint}
  \eprint{arXiv:gr-qc/1202.2788})

\bibitem{CWSIGNALS}
Prix R 2009 {\em {Neutron Stars and Pulsars}\/} (ed. W. Becker,
  Springer-Verlag)

\bibitem{Abbott:2009rfa}
Abbott B {\em et~al.\/} (The LIGO Scientific Collaboration and the Virgo
  Collaboration) 2010 {\em Astrophys. J.\/} {\bf 713} 671--685

\bibitem{Abbott:2008fx}
Abbott B {\em et~al.\/} (The LIGO Scientific Collaboration) 2008 {\em
  Astrophys. J.\/} {\bf 683} L45--L50

\bibitem{Abadie:2011md}
Abadie J {\em et~al.\/} (The LIGO Scientific Collaboration, the Virgo
  Collaboration) 2011 {\em Astrophys. J.\/} {\bf 737} 93

\bibitem{cw_moriond}
Palomba C, the LIGO Scientific~Collaboration and the Virgo~Collaboration 2011
  Searches for continuous gravitational wave signals and stochastic backgrounds
  in ligo and virgo data {\em 46th Rencontres De Moriond: Gravitational Waves
  And Experimental Gravity 20-27 Mar 2011, La Thuile, Aosta Valley, Italy\/}
  (\textit{Preprint} \eprint{arXiv:astro-ph.IM/1201.3176})

\bibitem{CLEANING}
Acernese F {\em et~al.\/} (Virgo Collaboration) 2009 {\em Class. Quantum
  Grav.\/} {\bf 26} 204002

\bibitem{NOEMI}
Accadia T {\em et~al.\/} (Virgo Collaboration) 2012 {\em JPCS\/} {\bf 363}
  012037

\bibitem{1742-6596-243-1-012010}
Coughlin M, the LIGO Scientific~Collaboration and the Virgo~Collaboration 2010
  {\em JPCS\/} {\bf 243} 012010

\bibitem{0264-9381-22-18-S33}
Acernese F {\em et~al.\/} (Virgo Collaboration) 2005 {\em Class. Quantum
  Grav.\/} {\bf 22} S1189

\bibitem{0264-9381-22-18-S18}
Acernese F {\em et~al.\/} (Virgo Collaboration) 2005 {\em Class. Quantum
  Grav.\/} {\bf 22} S1041

\bibitem{Grishchuk:1974ny}
Grishchuk L~P 1974 {\em Sov. Phys. - JETP\/} {\bf 40} 409

\bibitem{Kosowsky:1992rz}
Kosowsky A, Turner M and Watkins R 1992 {\em Phys. Rev. Lett.\/} {\bf 69}
  2026--2029 revised version

\bibitem{Caldwell:1991jj}
Caldwell R and Allen B 1992 {\em Phys. Rev. D\/} {\bf 45} 3447--3468 revised
  version

\bibitem{Ferrari:1998ut}
Ferrari V, Matarrese S and Schneider R 1999 {\em Monthly Notices of the Royal
  Astronomical Society\/} {\bf 303} 247

\bibitem{Regimbau:2005ey}
Regimbau T and de~Freitas~Pacheco J 2006 {\em Astron. Astrophys.\/} {\bf 447} 1

\bibitem{regimbau:2005tv}
Regimbau T and de~Freitas~Pacheco J 2006 {\em Astrophys. J.\/} {\bf 642}
  455--461

\bibitem{Christensen:1992wi}
Christensen N 1992 {\em Phys. Rev. D\/} {\bf 46} 5250--5266

\bibitem{Allen:1997ad}
Allen B and Romano J 1999 {\em Phys. Rev. D\/} {\bf 59} 102001

\bibitem{Abadie:2011fx}
Abadie J {\em et~al.\/} (LIGO Scientific Collaboration and Virgo Collaboration)
  2012 {\em Phys. Rev. D\/} {\bf 85}(12) 122001

\bibitem{1742-6596-228-1-012015}
Accadia T {\em et~al.\/} (Virgo Collaboration) 2010 {\em JPCS\/} {\bf 228}
  012015

\bibitem{0264-9381-28-2-025005}
Accadia T {\em et~al.\/} (Virgo Collaboration) 2011 {\em Class. Quantum
  Grav.\/} {\bf 28} 025005

\bibitem{SUPERATTENUATOR}
Acernese F {\em et~al.\/} (Virgo Collaboration) 2010 {\em Astroparticle
  Physics\/} {\bf 33} 182 -- 189

\bibitem{INJECTION}
Bondu F, Brillet A, Cleva F, Heitmann H, Loupias M, Man C, H T and the
  Virgo~Collaboration 2002 {\em Class. Quantum Grav.\/} {\bf 19} 1829

\bibitem{PhysRevA.79.053824}
Acernese F {\em et~al.\/} (Virgo Collaboration) 2009 {\em Phys. Rev. A\/} {\bf
  79}(5) 053824

\bibitem{PhysRevD.56.6085}
Vinet J~Y, Brisson V, Braccini S, Ferrante I, Pinard L, Bondu F and Tourni\'e E
  1997 {\em Phys. Rev. D\/} {\bf 56}(10) 6085--6095

\bibitem{springerlink:10.1007/BF00702605}
Drever R, Hall J, Kowalski F, Hough J, Ford G, Munley A and Ward H 1983 {\em
  Appl. Phys. B\/} {\bf 31}(2) 97--105

\bibitem{Acernese200829}
Acernese F {\em et~al.\/} (Virgo Collaboration) 2008 {\em Astroparticle
  Physics\/} {\bf 30} 29 -- 38

\bibitem{Accadia2011521}
Accadia T {\em et~al.\/} (Virgo Collaboration) 2011 {\em Astroparticle
  Physics\/} {\bf 34} 521 -- 527

\bibitem{NOISEBUDGET}
Acernese F {\em et~al.\/} (Virgo Collaboration) 2007 {Noise budget and noise
  hunting in Virgo} {\em {2007 Gravitational Waves and Experimental Gravity}\/}
  ed {Dumarchez, J, Tr\^an Thanh V\^an, J } pp 147--152

\bibitem{chatterjiThesis}
Chatterji S 2005 {\em {The search for gravitational wave bursts in data from
  the second LIGO science run}\/} Ph.D. thesis Massachusetts Institute of
  Technology \urlprefix\url{http://hdl.handle.net/1721.1/34388}

\bibitem{TCS}
Accadia T {\em et~al.\/} (Virgo Collaboration) 2010 {A Thermal Compensation
  System for the gravitational wave detector Virgo} ed {Damour, T, Jantzen, RT,
  Ruffini, R} pp 1652 -- 1656

\bibitem{ADV}
  \urlprefix\url{https://wwwcascina.virgo.infn.it/advirgo/}

\bibitem{0264-9381-27-8-084021}
Lorenzini M and the Virgo~Collaboration 2010 {\em Class. Quantum Grav.\/} {\bf
  27} 084021

\bibitem{GVAJENTE}
Vajente G 2008 {\em {Analysis of sensitivity and noise sources for the Virgo
  gravitational wave interferometer}\/} Ph.D. thesis Scuola Normale di Pisa

\bibitem{0264-9381-26-20-204007}
Leroy N, the LIGO Scientific~Collaboration and the Virgo~Collaboration 2009
  {\em Class. Quantum Grav.\/} {\bf 26} 204007

\bibitem{0264-9381-27-19-194012}
Robinet F, the LIGO Scientific~Collaboration and the Virgo~Collaboration 2010
  {\em Class. Quantum Grav.\/} {\bf 27} 194012

\bibitem{0264-9381-27-19-194010}
Christensen N, the LIGO Scientific~Collaboration and the Virgo~Collaboration
  2010 {\em Class. Quantum Grav.\/} {\bf 27} 194010

\bibitem{VDB}
\urlprefix\url{https://vdb.virgo.infn.it/main.php}

\bibitem{LINEDB}
\urlprefix\url{https://pub3.virgo.infn.it/MonitoringWeb/Noise/html/index.php?c%
allContent=408}

\bibitem{0264-9381-27-19-194013}
Buskulic D, the LIGO Scientific~Collaboration and the Virgo~Collaboration 2010
  {\em Class. Quantum Grav.\/} {\bf 27} 194013

\bibitem{0264-9381-27-19-194011}
Accadia T {\em et~al.\/} (Virgo Collaboration) 2010 {\em Class. Quantum
  Grav.\/} {\bf 27} 194011

\bibitem{0264-9381-28-23-235008}
{Coughlin, M and the LIGO Scientific Collaboration and the Virgo Collaboration}
  2011 {\em Class. Quantum Grav.\/} {\bf 28} 235008

\bibitem{pq_veto}
Ballinger T, the LIGO Scientific~Collaboration and the Virgo~Collaboration 2009
  {\em Class. Quantum Grav.\/} {\bf 26} 204003

\bibitem{Acernese2010131}
Acernese F {\em et~al.\/} (Virgo Collaboration) 2010 {\em Astroparticle
  Physics\/} {\bf 33} 131 -- 139

\bibitem{ALIGNMENT}
Accadia T {\em et~al.\/} (Virgo Collaboration) 2011 {\em Astroparticle
  Physics\/} {\bf 34} 327 -- 332

\bibitem{Bizouard:2007}
Bizouard M~A, Cavalier F, Christensen N, Clapson A and Hello P 2007 {Data
  quality and veto studies for the all-sky gravitational wave burst search in
  Virgo C7 Run data} Tech. Rep. VIR-0013D-07

\bibitem{0264-9381-25-18-184003}
Acernese F {\em et~al.\/} (Virgo Collaboration) 2008 {\em Class. Quantum
  Grav.\/} {\bf 25} 184003

\bibitem{kleinewelle}
Chatterji S {\em et~al.\/} 2004 {\em Class. Quantum Grav.\/} {\bf 21} S1809

\bibitem{Isogai:2010zz}
Isogai T, the LIGO Scientific~Collaboration and the Virgo~Collaboratio 2010
  {\em JPCS\/} {\bf 243} 012005

\bibitem{0264-9381-28-23-235005}
Smith J, Abbott T, Hirose E, Leroy N, MacLeod D, McIver J, Saulson P and
  Shawhan P 2011 {\em Class. Quantum Grav.\/} {\bf 28} 235005

\bibitem{LOGBOOK}
\urlprefix\url{{https://pub3.ego-gw.it/logbook}}

\bibitem{0264-9381-27-16-165023}
Slutsky J {\em et~al.\/} 2010 {\em Class. Quantum Grav.\/} {\bf 27} 165023

\bibitem{0264-9381-25-18-184006}
Gouaty R and the LIGO Scientific~Collaboration 2008 {\em Class. Quantum
  Grav.\/} {\bf 25} 184006

\bibitem{:2012na}
Abadie J {\em et~al.\/} (LIGO Scientific Collaboration and Virgo Collaboration)
  2012 {\em Phys. Rev. D\/} {\bf 85}(10) 102004

\bibitem{PhysRevD.68.102001}
Arnaud N, Barsuglia M, Bizouard M~A, Brisson V, Cavalier F, Davier M, Hello P,
  Kreckelbergh S and Porter E 2003 {\em Phys. Rev. D\/} {\bf 68}(10) 102001

\bibitem{PhysRevD.71.062001}
Allen B 2005 {\em Phys. Rev. D\/} {\bf 71}(6) 062001

\bibitem{Rodriguez:2008kt}
Rodriguez A 2008  (\textit{Preprint} \eprint{arXiv:gr-qc/0802.1376})

\bibitem{0264-9381-25-11-114029}
Klimenko S, Yakushin I, Mercer A and Mitselmakher G 2008 {\em Class. Quantum
  Grav.\/} {\bf 25} 114029

\bibitem{PhysRevD.72.122002}
Klimenko S, Mohanty S, Rakhmanov M and Mitselmakher G 2005 {\em Phys. Rev. D\/}
  {\bf 72}(12) 122002

\bibitem{EM_FOLLOWUP}
Abadie J {\em et~al.\/} (The LIGO Scientific Collaboration, the Virgo
  Collaboration) 2012 {\em Astron. Astrophys.\/} {\bf 539} 184006

\bibitem{0264-9381-22-18-S34}
Astone P, Frasca S and Palomba C 2005 {\em Class. Quantum Grav.\/} {\bf 22}
  S1197

\bibitem{0264-9381-22-18-S39}
Palomba C, Astone P and Frasca S 2005 {\em Class. Quantum Grav.\/} {\bf 22}
  S1255

\bibitem{0264-9381-24-19-S12}
Acernese F {\em et~al.\/} (Virgo Collaboration) 2007 {\em Class. Quantum
  Grav.\/} {\bf 24} S491

\bibitem{0264-9381-24-19-S07}
Acernese F {\em et~al.\/} (Virgo Collaboration) 2007 {\em Class. Quantum
  Grav.\/} {\bf 24} S433

\bibitem{0264-9381-24-19-S32}
Mukherjee S and the LIGO Scientific~Collaboration 2007 {\em Class. Quantum
  Grav.\/} {\bf 24} S701

\bibitem{0264-9381-28-15-155001}
Stroeer A, Blackburn L and Camp J 2011 {\em Class. Quantum Grav.\/} {\bf 28}
  155001

\bibitem{Costa:2011jc}
Costa C~A and Torres C~V 2011  (\textit{Preprint}
  \eprint{arXiv:physics.data-an/1111.4516})

\bibitem{MNR:MNR18818}
Pitkin M 2011 {\em Monthly Notices of the Royal Astronomical Society\/} {\bf
  415} 1849--1863

\bibitem{0264-9381-27-17-173001}
Abadie J {\em et~al.\/} 2010 {\em Class. Quantum Grav.\/} {\bf 27} 173001

\end{thebibliography}
